\pdfoutput=1
\documentclass[11pt,a4paper]{article}

\usepackage[margin=1in]{geometry}
\usepackage[T1]{fontenc}
\usepackage{amsmath,amssymb,amsthm}
\usepackage{booktabs}
\usepackage{enumitem}
\usepackage{tikz}
\usepackage{pgfplots}
\pgfplotsset{compat=1.17}
\usetikzlibrary{arrows.meta,decorations.pathmorphing,positioning}
\usepackage[most]{tcolorbox}
\usepackage{listings}
\usepackage[colorlinks=true,linkcolor=black,citecolor=black,urlcolor=blue]{hyperref}

\newtheorem{theorem}{Theorem}
\newtheorem{corollary}{Corollary}
\theoremstyle{definition}
\newtheorem{example}{Example}

\newtcolorbox{translation}[1][]{
  colback=black!4, colframe=black!45, boxrule=0.4pt, arc=2pt,
  left=6pt,right=6pt,top=5pt,bottom=5pt,
  fonttitle=\bfseries\small, title={Engineering remark}, #1}

\newtcolorbox{warnbox}[1][]{
  colback=black!2, colframe=black!60, boxrule=0.6pt, arc=0pt,
  left=6pt,right=6pt,top=5pt,bottom=5pt,
  fonttitle=\bfseries\small, title={Notational comment}, #1}

\newcommand{\Ab}{\bar{A}}
\newcommand{\Aone}{A_{1,1}}
\newcommand{\Atwo}{A_{2,1}}

\title{Symmetry and the Form of Nonlinear Behavioral Models\\[4pt]
\large A tutorial for microwave engineers}
\author{N.~B.~Tufillaro\thanks{Aqualytics, Corvallis, Oregon. \texttt{nick@aqualytics.eco}}}
\date{Tutorial, revised 27 November 2026\\[4pt]
\normalsize\textit{Companion to the paper ``Time invariance, circle symmetry, and the
completeness of the Cardiff behavioral model''~\cite{paper1} and to the MATLAB toolkit
of Appendix~\ref{app:matlab}~\cite{toolkit}.}}

\begin{document}
\maketitle

\begin{abstract}
\noindent
Behavioral models of nonlinear microwave devices --- the Cardiff model, X-parameters, the
higher-order describing functions of the mechanical-systems literature --- all share a
functional form. This tutorial derives that form from time invariance alone. It develops the
consequences of that derivation using only the harmonic-balance description of a
single-tone periodic steady state, in which each port waveform is a finite set of
harmonic phasors, and uses them to give the Cardiff model's three
exponents physical interpretations. In the monomial rewriting the magnitude exponent $m$ is the order of the device's
load-side nonlinearity; the phase exponent $n$ is set by the drive-side harmonic; and the conjugate
index $r$ obeys $r_{\max}=\lfloor K/2\rfloor$, where $K$ is the degree of the load-side
nonlinearity. The familiar restriction $r\le1$ is therefore a hypothesis about the
device, exact whenever $K\le3$; for a device closed through its load the bound is replaced
by a decay in $r$. The final sections
show that a tailored A-pull measurement displays this decomposition directly: each spectral
cluster's half-width is the lowest order of nonlinearity able to produce it. The statements
are illustrated with a published measurement and two simulations; they have not been
tested against a measurement in which $K$ is known independently. The tutorial is
pedagogical and is meant as a gentler introduction to the results treated in the companion
paper ``Time invariance, circle symmetry, and the completeness of the Cardiff behavioral
model''~\cite{paper1}, which states and proves the theorems in full.
\end{abstract}

\tableofcontents
\bigskip

\section{Introduction}

A behavioral model is usually presented in a fixed sequence: a functional form is written
down and motivated by physical reasoning or by experience, the form is fitted, and the
quality of the fit is reported. The Cardiff model was introduced in this
way~\cite{qi2009,woodington2010}, and its authors, like those of X-parameters and of the
describing-function models, gave time invariance as the reason for the form. Woodington
\emph{et al.} observe that the device ``responds only to the magnitudes of the signals
\ldots\ and the relative phase difference''~\cite{woodington2010}, and Tasker and Benedikt
write the general two-port response at harmonic $h$ as the $h$th power of the drive phase
times an arbitrary function of two
magnitudes and a relative phase~\cite{tasker2011,tasker2009dml}; Root \emph{et al.} call
the phase factor and the magnitude-only coefficients ``a necessary consequence of the
assumed time invariance of the underlying system''~\cite{root2005phd}, and the
X-parameter book derives the form by fixing the phase reference~\cite{root2013}; and Nuij
\emph{et al.} build the same phase factor into the describing-function models as a
``virtual harmonics generator''~\cite{nuij2006}. In the power-amplifier predistortion
literature, Enzinger proved the same symmetry for a general time-invariant operator with
memory, for one input, and named it ``phase homogeneity''~\cite{enzinger2018}.

This tutorial presents the time-invariance argument at an elementary level and in a form that covers
the three models together. Once it is accepted that nothing physical depends on where the
clock is started, the functional form of every behavioral model on a harmonic grid is
determined up to a set of arbitrary smooth coefficient functions: the structure of the
model is forced, and only the coefficient functions remain to be fitted. This tutorial
clarifies the connection between some classical mathematics and the form and properties
of the behavioral model, in particular:

\begin{enumerate}[label=(\arabic*),leftmargin=2em]
\item \textbf{The form is complete.} Every time-invariant response of a two-port in
      single-tone periodic steady state has this form, with or without memory. If a
      single-tone continuous-wave (CW) fit leaves a residual that stays at the same level
      as terms are added, what is missing is something outside the form: another tone,
      another port, harmonic injection, or a measurement that is not in
      steady state. The statement says which terms can occur. It gives no estimate of how
      many are needed for a given accuracy.
\item \textbf{The exponents have physical interpretations.} The Cardiff indices $m$, $n$
      and $r$ are not only labels attached to basis functions. Written as monomials they
      are counters: $m$ counts how many times the load-side wave entered, $n$ counts the
      harmonic the drive side supplied, and $r$ counts conjugate pairs.
\item \textbf{The restriction $r\le1$ can be interpreted as a property of the device.}
      It is exact when the device's load-side nonlinearity is at most cubic, and it has a
      measurable signature.
\end{enumerate}

\paragraph{Prerequisites.} You need phasors, harmonic balance and the binomial theorem
--- every term is defined
where it is first used, and the glossary in Section~\ref{sec:glossary} collects them with
engineering interpretations. The mathematics used here is classical. Section~\ref{sec:proof} contains the one result
specific to the Cardiff model. The material overlaps largely
with the companion paper ``Time invariance, circle symmetry, and the completeness of the
Cardiff behavioral model,'' which states and proves the theorems in full; this tutorial
starts at a more elementary level and is meant as a gentler introduction to the results
treated in~\cite{paper1}. Memory, which this tutorial leaves aside, is the
subject of a second companion paper~\cite{paper2}. The tutorial can also be read as an
elementary introduction, for engineers, to the representation theory of the circle group
and to classical invariant theory, introduced through a single application.

\paragraph{Route through the paper.} Sections~\ref{sec:warmup}--\ref{sec:equivariance}
build the symmetry argument.
Sections~\ref{sec:monomials}--\ref{sec:corollaries} apply it to the Cardiff exponents.
Section~\ref{sec:loaded} deals with the loaded device, where Theorem~\ref{thm:rmax} no
longer applies and a decay rule takes its place. Section~\ref{sec:apull} connects the
results to a published measurement. A reader who wants only the result can
read Section~\ref{sec:proof} and Appendix~\ref{app:noalgebra}, which is the same argument
without using group theory.

\section{A symmetry argument}
\label{sec:warmup}

Consider a memoryless nonlinearity with odd symmetry,
\begin{equation}
i = F(v), \qquad F(-v) = -F(v),
\end{equation}
and drive it with a single tone $v(t) = V\cos\omega_0 t$. The output contains only odd
harmonics, and the reason is a symmetry argument. Shift time by half a period, $t \to t + T/2$. The drive
changes sign, $v \to -v$. By odd symmetry the response also changes sign, $i \to -i$. But
the $k$-th harmonic of the response picks up a factor $e^{jk\omega_0 T/2} = e^{jk\pi} =
(-1)^k$ under that same shift. So every harmonic must satisfy $(-1)^k I_k = -I_k$, which
forces $I_k = 0$ for even $k$.

Notice what that argument did \emph{not} do. It did not expand $F$, it did not assume a
polynomial, it did not evaluate an integral, and it did not depend on the form of $F$.
It ruled out half the possible answers using only a symmetry. That is the entire
method of this tutorial; in the sections that follow, only the symmetry group changes.

\paragraph{Naming the pieces.} Three ingredients recur.

\begin{itemize}[leftmargin=1.4em]
\item A \textbf{group} of transformations that leave the physics unchanged. Here it is the
      two-element set $\{\text{identity}, \text{shift by } T/2\}$, where the identity is the
      transformation that does nothing. ``Group'' means only that
      the transformations compose, there is an identity, and every transformation
      can be undone --- which is true of any symmetry. This
      particular group is called $\mathbb{Z}_2$.
\item An \textbf{action}: a rule saying how each signal transforms. Here the drive picks up
      a factor $-1$ and harmonic $k$ picks up $(-1)^k$. Different quantities transform
      differently under the same symmetry, and keeping track of that is most of the work.
\item A conclusion about which terms can \textbf{survive}. A quantity unchanged by every
      group element is called \textbf{invariant}; one that transforms in a prescribed way
      is called \textbf{equivariant}.
\end{itemize}

\paragraph{An elementary example.} Take the sign flip $v\to-v$ of the argument above.
The square $v^{2}$ is an invariant: it is unchanged by the flip. The cube $v^{3}$ is an
equivariant: it changes sign with $v$, which is the prescribed way for any quantity that
must behave like the drive. Among the harmonics of the response, the fundamental $I_1$
picks up $-1$ under the half-period shift and so is equivariant in that same sense, while
the second harmonic $I_2$ picks up $+1$ and so is invariant; the response of an odd
nonlinearity must change sign with the drive, so its invariant harmonics are zero. The
same two notions appear for rotations of the plane: the length of a vector is invariant,
and the vector itself is equivariant, since it rotates with the rotation. That is the form
the notions take in Section~\ref{sec:invariants}, where the magnitude of a phasor is the
invariant and the phasor is the equivariant.

\paragraph{A second example, from microwave practice.} Drive a mixer with two tones at
$\omega_1$ and $\omega_2$. The output appears only at frequencies $m\omega_1 + n\omega_2$
with $m, n$ integers --- never at, say, $\omega_1/2$ or $\sqrt{2}\,\omega_1$.\footnote{Unless
the symmetry breaks. A parametric divider produces $\omega_1/2$ precisely by
\emph{spontaneously} breaking this phase symmetry: the half-frequency output has two
equally good phases and the circuit picks one. Period doubling in nonlinear dynamics is the
same phenomenon. So the argument is not that subharmonics are impossible; it is that they
are a broken-symmetry state, and the symmetric steady state assumed throughout this
tutorial excludes them.} The reason is the same. The phase of each input may be shifted independently, and each output component
must track the two shifts coherently; only integer combinations can. The symmetry group
this time is two independent circles of phase, a \emph{two-torus}, and the integers $m$ and
$n$ are exactly the bookkeeping of how a given output component responds to them.

That structure returns in Sections~\ref{sec:monomials}--\ref{sec:corollaries}, where the Cardiff exponents are its $m$ and $n$.

\begin{translation}
A symmetry is a statement that some experimental choice was arbitrary --- the origin of
time, the absolute phase of a generator. Anything you
can measure must be independent of that choice. That independence is a constraint, and
constraints eliminate candidate model terms before any fitting.
\end{translation}

\section{The symmetry behind behavioral models}
\label{sec:symmetry}

\subsection{Where the clock is started}

Consider a device in periodic steady state at fundamental $\omega_0$. Every port voltage,
current and traveling wave is described by a set of harmonic phasors: $A_{p,k}$ for the
incident wave at port $p$, harmonic $k$, and so on. The time-domain waveform is
\begin{equation}
a_p(t) = \operatorname{Re}\sum_{k\ge0} A_{p,k}\, e^{jk\omega_0 t}.
\end{equation}
The scattered wave at port $p$ is written in the same way, with harmonic index $h$,
$b_p(t)=\operatorname{Re}\sum_{h\ge0}B_{p,h}\,e^{jh\omega_0 t}$; $k$ is used for input
harmonics and $h$ for output harmonics throughout, so that the two can be discussed
together.

Now ask what happens if the clock had been started at time $\tau$ instead of at $0$. A
later time $t=t'+\tau$ is read by the shifted clock as $t'$. The waveform is physically
identical --- the same experiment, the same device --- but the phasors that describe the
same waveform in terms of $t'$ are multiplied by phase factors that depend on the harmonic
index:
\begin{equation}
A_{p,k} \;\longmapsto\; A_{p,k}\, e^{jk\omega_0\tau}.
\label{eq:action}
\end{equation}
Each harmonic rotates, and the $k$-th harmonic rotates $k$ times as fast as the
fundamental. Write $\theta = \omega_0\tau$; since a shift of one full period leaves every
phasor unchanged, $\theta$ is an angle on a circle, $\theta\in[0,2\pi)$.

Equation~\eqref{eq:action} is the key physical assumption used in Sections~\ref{sec:invariants}
and~\ref{sec:equivariance}; Section~\ref{sec:proof} adds one more, that the device is
memoryless with a polynomial dependence of degree $K$ on the load-side voltage.

\begin{figure}[htbp]
\centering
\begin{tikzpicture}[scale=1.0]
  \foreach \k/\x in {1/0, 2/4.2, 3/8.4}{
    \begin{scope}[shift={(\x,0)}]
      \draw[black!25] (-1.5,0)--(1.5,0);
      \draw[black!25] (0,-1.5)--(0,1.5);
      \draw[black!55] (0,0) circle (1.2);
      \draw[-{Stealth[length=2.5mm]},thick] (0,0)--(25:1.2);
      \node at (25:1.45) {\footnotesize $A_{p,\k}$};
      \pgfmathsetmacro\ang{25+\k*38}
      \draw[-{Stealth[length=2.5mm]},thick,black!55,dashed] (0,0)--(\ang:1.2);
      \draw[-{Stealth[length=2mm]},black!70] (25:0.72) arc (25:\ang:0.72);
      \node at (0,-1.85) {\footnotesize $k=\k$: rotates by $\k\theta$};
    \end{scope}}
\end{tikzpicture}
\caption{Starting the clock at time $\tau$ rotates the $k$-th harmonic phasor by
$k\theta$, $\theta=\omega_0\tau$. All harmonics rotate together, but at rates set by their
harmonic index. The integer $k$ is called the \emph{weight} of that harmonic.}
\label{fig:weights}
\end{figure}

\subsection{This set of transformations is a circle group}

The transformations \eqref{eq:action}, one for each $\theta$, compose by adding angles:
doing $\theta_1$ then $\theta_2$ is the same as doing $\theta_1+\theta_2$. Doing $\theta=0$
is the identity, and $-\theta$ undoes $\theta$. So they form a group, and as a set that group
\emph{is} a circle. It has several names, all for the same object:
\[
U(1) \;=\; SO(2) \;=\; S^1 \;=\; \{e^{j\theta}\}.
\]
The names come from four ways of looking at the same circle: $U(1)$ is the group of
unitary $1\times1$ matrices, that is, of complex numbers of modulus one, the pure phase
factors; $SO(2)$ is the group of rotations of the plane, the same phase shifts seen as
rotations of a phasor diagram; $S^1$ is the circle itself, as a set of points; and
$\{e^{j\theta}\}$ is the circle as an engineer writes it. The distinctions matter only
when the names appear in the theorems quoted later, where the usual one is $U(1)$; for
this tutorial the reader can take all four to mean the phase knob on a signal generator.

It differs from the $\mathbb{Z}_2$ of Section~\ref{sec:warmup} in one important way. It has
a \emph{continuous} parameter, and composition is a smooth function of that parameter. A
group with that property is called a \textbf{Lie group}. The property is what lets one use calculus on the group, and it is the hypothesis in the
theorems quoted later.

The assignment of the factor $e^{jk\theta}$ to harmonic $k$ is called a
\textbf{representation} of the group, and the integer $k$ is called the \textbf{weight}
of that quantity.\footnote{A physicist would call the weight the \emph{charge} of the
quantity under the $U(1)$ symmetry, by analogy with electric charge, which is the weight
under the phase rotations of electromagnetism. In that vocabulary the rule of
Section~\ref{sec:invariants}, that a product is invariant when its total weight is zero,
is charge conservation: only neutral combinations are observable; and the choice of
carrier in Section~\ref{sec:equivariance} is a choice of gauge. The engineering content
is unchanged by the names.} Weight is a property of how a quantity transforms, not of
its size: every fundamental-frequency phasor has weight one, whatever its magnitude, and
normalizing a phasor to unit magnitude, $A/|A|$, leaves its weight unchanged. Conjugation
reverses the sign of the weight, since $\bar A$ picks up $e^{-jk\theta}$. Equivalently,
the weight is a winding number: as $\theta$ goes once around the circle, a quantity of
weight $k$ goes $k$ times around the unit circle, in the negative sense if $k<0$.
Figure~\ref{fig:weights} is a picture of the weights $1$, $2$, $3$.

\begin{translation}
$U(1)$ is the phase knob on a signal generator, taken as a mathematical object. The
statement ``harmonic $k$ has weight $k$'' is the statement that turning one phase knob on
the fundamental rotates the $k$-th harmonic $k$ times as fast, which is what an
oscilloscope shows when a delay line is changed.
\end{translation}

\subsection{More circles: extra tones, and why reference planes are different}

For a single tone and its harmonics the time origin is the only freedom of this kind. If
the device is driven by several incommensurate tones, each tone carries its own phase
origin, another circle each. Two incommensurate tones give a product of two circles,
\[
T^{2} \;=\; U(1)\times U(1),
\]
a \emph{two-torus}, and several tones give a torus with one circle per tone. The mixer
example of Section~\ref{sec:warmup} is the two-torus case. Commensurate tones share a
fundamental and reduce to a single circle with the appropriate integer weights. The
two-torus case exists in
measurement practice: the multi-tone multi-harmonic scattering (M$^2$S) parameters of
Gibiino \emph{et al.}~\cite{gibiino2020}
drive two incommensurate tones, index the response by the integer pair $(m,n)$, and have
coefficients that depend on the two magnitudes alone, in the authors' words because
``no phase dependency is involved due to time-invariance.''

Moving a reference plane is a freedom of a different kind. Moving the reference plane at
port $p$ along a dispersionless line shifts the phase of every wave at that port, incident
and reflected in opposite senses, and harmonic $k$ by $k$ times the fundamental's
shift~\cite{kurokawa1965}. The description of the device changes with it, as
$S$-parameters change by phase factors: the map from incident to scattered waves is
replaced by a different map. That places no constraint on the map of one device, so a
reference-plane move is a change of description and is not a symmetry in the sense used
here. The rest of this tutorial uses only the single circle of the time origin, one tone
together with its harmonics.

\section{What survives the rotation: invariants}
\label{sec:invariants}

\subsection{The question}

A device response is a function of the incident waves. Suppose it is a function of the
magnitudes and phases of $\Aone$ and $\Atwo$ (drive-side and load-side fundamental
incident waves). The symmetry says that if both are rotated --- $\Aone \to \Aone
e^{j\theta}$, $\Atwo \to \Atwo e^{j\theta}$, since both have weight one --- nothing physical
can change. Which functions of $\Aone$ and $\Atwo$ have that property?

Functions unchanged by every group element are called \textbf{invariants}.

\subsection{The answer for a circle}

Two kinds of quantity survive (i.e.\ are invariant).

\paragraph{Magnitudes.} $|\Aone|$ and $|\Atwo|$ are invariant: rotation does not
change length. Figure~\ref{fig:orbit} shows the orbit. As $\theta$ runs over the circle, a
phasor traces a circle of constant radius; that circle is called the \textbf{orbit} of the
phasor, and the radius is the invariant that labels the orbit.

\paragraph{Weight-balanced products.} Individual phases do not survive, but differences do.
Consider
\[
\Atwo \Ab_{1,1}\;\longmapsto\; \Atwo e^{j\theta}\,\Ab_{1,1}e^{-j\theta} = \Atwo\Ab_{1,1}.
\]
The two phase factors cancel because the weights cancel: $+1$ from $\Atwo$, $-1$ from the
conjugate of $\Aone$. In general a monomial $\prod A_{p,k}^{\alpha}\Ab_{p,k}^{\beta}$ is
invariant precisely when its total weight $\sum k(\alpha - \beta)$ is zero. That is a
single linear condition on the exponents.

Dividing out the magnitudes leaves the object the Cardiff literature calls the relative
phase:
\begin{equation}
Q \;\equiv\; \frac{\Atwo/|\Atwo|}{\Aone/|\Aone|} \;=\; e^{j(\phi_{2,1}-\phi_{1,1})},
\qquad \phi_{p,k}\equiv\angle A_{p,k}.
\end{equation}

\begin{figure}[htbp]
\centering
\begin{tikzpicture}[scale=1.1]
  \begin{scope}
  \draw[black!25] (-1.7,0)--(1.7,0) node[right,black!60,font=\footnotesize]{$\operatorname{Re}$};
  \draw[black!25] (0,-1.7)--(0,1.7) node[above,black!60,font=\footnotesize]{$\operatorname{Im}$};
  \draw[black!65,thick] (0,0) circle (1.25);
  \draw[-{Stealth[length=2.5mm]},very thick] (0,0)--(35:1.25);
  \node at (35:1.5) {\footnotesize $A$};
  \draw[-{Stealth[length=2mm]},black!60] (100:1.25) arc (100:160:1.25);
  \node[black!60,font=\footnotesize] at (130:1.62) {orbit};
  \node[font=\footnotesize,black!70,anchor=north,rotate=35] at (35:0.62) {$|A|$};
  \node[font=\footnotesize,align=center] at (0,-2.15)
     {radius is invariant\\ position on the circle is not};
  \end{scope}
  \begin{scope}[shift={(5.2,0)}]
  \draw[black!25] (-1.7,0)--(1.7,0);
  \draw[black!25] (0,-1.7)--(0,1.7);
  \draw[-{Stealth[length=2.5mm]},very thick] (0,0)--(20:1.3) node[right,font=\footnotesize]{$\Aone$};
  \draw[-{Stealth[length=2.5mm]},very thick] (0,0)--(78:0.95) node[above,font=\footnotesize]{$\Atwo$};
  \draw[-{Stealth[length=2.5mm]},thick,black!45,dashed] (0,0)--(65:1.3);
  \draw[-{Stealth[length=2.5mm]},thick,black!45,dashed] (0,0)--(123:0.95);
  \draw[black!70] (20:0.45) arc (20:78:0.45);
  \node[font=\footnotesize] at (49:0.68) {$\Delta\phi$};
  \node[font=\footnotesize,align=center] at (0,-2.15)
     {rotate both: $\Delta\phi$ survives};
  \end{scope}
\end{tikzpicture}
\caption{Left: the orbit of a phasor under the circle group, and the invariant that labels
it. Right: absolute phases are not invariant, but the difference between two phasors of
weight one (two fundamentals) is. These two facts generate every invariant there is.}
\label{fig:orbit}
\end{figure}

\subsection{Why that is the complete list}

It is not obvious that magnitudes and relative phases exhaust the possibilities --- perhaps
some other function of the phases is invariant without being built from differences.
Classical invariant theory says no. For a compact group (and a circle is compact), the
invariants are generated by finitely many basic ones, a result usually attributed to
Hilbert and Weyl. Any invariant \emph{polynomial} is a polynomial in the generators.

There is also a constructive way to see the result, one that engineers already use. Take
any function of the phasors and average it over the phase angle $\theta$, from $0$ to
$2\pi$. What survives the average is invariant by construction, and every invariant is
its own average, so this ``phase averaging'' is a projection onto the invariants. In
invariant theory it is called the Reynolds operator; for a circle it is
integration over one period of the drive phase. That the group is \emph{compact} --- a
circle has finite length --- is what makes the average well defined, and it is the
hypothesis on which everything in this section rests. Finite groups (average over the
elements), circles and tori (average over the angles) all qualify.

For devices that is not enough, because measured device responses are
smooth but not polynomial. A theorem of Schwarz~\cite{schwarz1975} extends the result
beyond polynomials: for a compact Lie group, every smooth invariant function is a
smooth function of the same finite set of generators. Nothing is lost in passing from
polynomial to smooth functions. Where both waves are nonzero, though, the theorem is more than
is needed: there the magnitudes and the relative phase are ordinary smooth coordinates,
and the X-parameter book's division of the response by the drive
phase~\cite{root2013} is itself the proof. Schwarz's theorem is needed only at
zero load wave, where the relative phase is undefined, and that is the case treated next.

\begin{translation}
``Generated by'' means: write your model in terms of $|\Aone|$, $|\Atwo|$ and $Q$, and you
are not leaving anything out. Not approximately --- exactly, and for smooth responses, not
just polynomial ones. Any term you might add on physical grounds is already a function of
these three.
\end{translation}

The relative phase $Q$ is a ratio, and it is undefined
when the load wave vanishes. A response that is smooth \emph{at} $|\Atwo|=0$ cannot
therefore depend on $Q^{n}$ alone; it must carry enough powers of $|\Atwo|$ to cancel the
singularity. The term $|\Atwo|^{m}Q^{n}$ is a polynomial in $\Atwo$ and $\bar{A}_{2,1}$
--- hence smooth at the origin --- exactly when $m\ge|n|$ and $m-|n|$ is even, that is,
when $m=|n|+2r$ for some integer $r\ge0$. That is the Cardiff index relation: the smoothness condition at zero load wave, which in
the rewriting of~\cite{tian2025} (Section~\ref{sec:monomials}) is the condition that each
term be a polynomial in $\Atwo$ and its conjugate. One caution goes with it. The series in
these terms is the Taylor expansion of the response at zero load wave. It equals the
response where the response is analytic; for a response that is smooth without being
analytic it is an asymptotic expansion, with a remainder. The same distinction is drawn, for one input, in the baseband
modeling of power amplifiers, between the odd-order monomials $x^{k+1}\bar x^{k}$ and the
even-order terms $|x|^{2k+1}x$, which satisfy the symmetry but are not polynomial in $x$
and $\bar x$~\cite{enzinger2018,ding2004}.

\begin{warnbox}
In the Cardiff literature $\angle A$ does not denote the angle of $A$ --- it denotes the
\emph{unit phasor} $e^{j\angle A}$. So the factor written $(\angle \Atwo / \angle
\Aone)^{n}$ means $e^{jn(\phi_{2,1}-\phi_{1,1})} = Q^n$, not a ratio of two real angles
raised to a power.
\end{warnbox}

\subsection{An aside: counting the invariants}

To know how many independent terms exist at each degree --- useful when
budgeting parameters --- there is a generating function for it, the Molien series. For a
circle acting with weights $w_1,\dots,w_N$ (and on the conjugates with $-w_i$), the number
of invariants of degree $d$ is the coefficient extracted from
\[
M(t) = \left[\lambda^{0}\right]\prod_{i}\frac{1}{(1-t\lambda^{w_i})(1-t\lambda^{-w_i})},
\]
and replacing $\lambda^0$ by $\lambda^h$ counts the weight-$h$ equivariants instead. For
two variables of weight one and their conjugates the answer is $(1+t^2)/(1-t^2)^3$, and the
shape of that fraction is itself informative. The denominator says there are three
algebraically independent invariants of degree two --- $|\Aone|^2$, $|\Atwo|^2$ and
$\operatorname{Re}(\Atwo\Ab_{1,1})$ --- in which one may form arbitrary polynomials;
the numerator says there is one more, $\operatorname{Im}(\Atwo\Ab_{1,1})$, which
appears at most linearly, because its square is already a polynomial in the other three.
Every invariant is uniquely $p + q\,\operatorname{Im}(\Atwo\Ab_{1,1})$ with $p,q$
polynomials in the first three. That is the general pattern (the Hironaka decomposition
into primary and secondary invariants), and a short account of it is a set of lectures on invariant theory for
physicists~\cite{demellokoch2026}, whose Sections~2.3--2.8 are entirely
field-independent. The finiteness theorem behind all of this is in Weyl's own 1936
lectures~\cite{weyl1936}. The accompanying MATLAB toolkit computes the series
(\texttt{xp\_molien}, \texttt{xp\_series}); Appendix~\ref{app:matlab} lists the routines.

\section{What the output must do: equivariance}
\label{sec:equivariance}

\subsection{The output is not invariant}

The device \emph{response} (the scattered wave) is not an invariant. $B_{p,h}$ is itself a harmonic-$h$ phasor, so under a shift of the
time origin it must rotate at weight $h$:
\begin{equation}
B_{p,h} \;\longmapsto\; B_{p,h}\, e^{jh\theta}.
\end{equation}
A map that transforms this way --- inputs rotate, output rotates by a prescribed weight ---
is called \textbf{equivariant}. Invariance is the special case $h=0$.

\subsection{Factoring an equivariant}

Once the invariants are known, the equivariants follow. Pick any convenient
quantity of weight $h$; call it the \emph{carrier}. Divide the output by it. The quotient
has weight zero, so it is an invariant, so by Section~\ref{sec:invariants} it is a smooth
function of the generators. Multiplying the quotient by the carrier again recovers the
output in factored form:
\begin{equation}
\boxed{\;\text{equivariant of weight } h \;=\;
(\text{carrier of weight } h)\times(\text{arbitrary invariant function}).\;}
\end{equation}

The natural carrier is the drive-side phase itself,
\begin{equation}
P \equiv e^{j\phi_{1,1}} = \frac{\Aone}{|\Aone|},
\end{equation}
which has weight one, so $P^h$ has weight $h$. Substituting gives
\begin{equation}
B_{p,h} \;=\; P^{h}\sum_{n} f_{p,h,n}\!\left(|\Aone|,|\Atwo|\right)\,Q^{\,n},
\qquad Q = \frac{\angle\Atwo}{\angle\Aone},
\label{eq:complete}
\end{equation}
with the $f_{p,h,n}$ \emph{arbitrary smooth functions} of two real variables. (The
hypothesis is that the response depends smoothly on $\Aone$ and $\Atwo$ away from zero;
the Cardiff form with $m=|n|+2r$ then follows from smoothness at $\Atwo=0$, as in
Section~\ref{sec:invariants}.)

Equation~\eqref{eq:complete} is the general solution, not an approximation, a
truncation or a modeling assumption. It carries no truncation estimate: it says which
terms can occur and says nothing about how many are needed. Every
time-invariant two-port response in single-tone periodic steady state has this form, and
the index $n$ is canonical rather than chosen: the sum over $n$ is the Fourier series in
the relative phase, so $n$ is a weight, a property of the group, not of the modeler.

Note what was \emph{not} assumed: memorylessness. A device with thermal or trapping
memory, held in CW steady state, still obeys \eqref{eq:complete}, because the only
physical hypothesis was that the response is unchanged when the clock is re-zeroed. What memory changes is the
behavior under modulation, where the coefficient functions acquire a dependence on the
history of the envelope, the slowly varying amplitude of the modulated drive, that no CW
measurement can detect. That is the same conclusion the
X-parameter literature reaches~\cite{root2013}: CW X-parameters are complete for CW, and
memory is a separate, dynamic extension. The dynamic gain model of Verspecht \emph{et
al.}~\cite{verspecht2023dg} imposes the same structure at the envelope level.\footnote{In that model the response is
an operator invariant under rotation of the phase of the RF tone, multiplied by the
envelope, with memory terms that vanish on a constant envelope.} The memoryless hypothesis is used later, in
Section~\ref{sec:proof}, because a polynomial degree in the load-side voltage is a property
of a static map. Equation~\eqref{eq:complete} is the form Tasker and Benedikt give for
fundamental load-pull, $b_{n,h}=P_1^{h}\,g(|a_{1,1}|,|a_{2,1}|,Q_1/P_1)$ at fixed bias,
with $g$ expanded in powers of the periodic phase vector $Q_1/P_1$ and coefficients that
are functions of the two magnitudes~\cite{tasker2011}; it appears earlier in Tasker's
2009 lecture~\cite{tasker2009dml}. It is also the form that the 2010 Cardiff
paper~\cite{woodington2010} fitted. What the symmetry argument adds is the statement,
with its proof, that the form is exhaustive for smooth responses.

\subsection{Gauge fixing}

Choosing $P$ as the carrier is a convention. The weight could equally be carried on
$\Atwo$'s phase, or on a fixed external reference. Different choices give different
coefficient functions $f$ describing the same physical device --- the components change,
the object does not. Physicists call this a choice of \emph{gauge}; microwave engineers make
the same choice every time they pick a reference plane, and moving the reference plane
changes the $S$-parameters by phase factors in exactly the way a change of carrier changes
the coefficient functions $f$.

The factor $P^h$ in \eqref{eq:complete} is the phase normalization
$P^{k}$ of X-parameters~\cite{root2013}, the ``virtual harmonics generator'' of the
higher-order sinusoidal describing function literature~\cite{nuij2006}, and the
phase-vector construction of the Cardiff model: $P^h$ is the gauge chosen in all three modeling frameworks.

\begin{translation}
Fixing the gauge is normalizing out the absolute phase, which is what you do anyway before
comparing two measurements. The mathematical content is that after you normalize, what is
left --- the coefficient functions $f_{p,h,n}$ --- is device data, independent of the
phase reference.
\end{translation}

\section{Writing the exponents as monomials}
\label{sec:monomials}

\subsection{From $(m,n,r)$ to $(a,b)$}

Expanding the coefficient functions in \eqref{eq:complete} in powers of $|\Atwo|$ gives
the traveling-wave form of the Cardiff model at fixed drive, with only the fundamental
load wave $\Atwo$ varied and no harmonic injection~\cite{woodington2010,azad2022}:
\begin{equation}
B_{p,h} \;=\; (\angle\Aone)^{h}\sum_{r}\sum_{n} C_{p,h,m,n}\,|\Atwo|^{m}
\left(\frac{\angle\Atwo}{\angle\Aone}\right)^{\!n},
\qquad m=|n|+2r .
\label{eq:cardiff}
\end{equation}
The Cardiff literature writes these coefficients $K_{p,h,m,n}$; they are written $C$ here
because $K$ is used below for the degree of the load-side nonlinearity.

The key step of the analysis is a change of variables, which Tian \emph{et al.}\ have
already made: they rewrite the Cardiff model as a polynomial in $\Atwo$ and its
conjugate, with general term $\binom{m}{k}(\Atwo)^{m-k}(\bar A_{2,1})^{k}$, ``where the
value of $n$ can now be calculated with $n=m-2k$, the value of $m$ equals to the
polynomial order number''~\cite{tian2025}. Write $A\equiv\Atwo$ and express a single term
as a monomial in $A$ and its conjugate:
\begin{equation}
|A|^{m}\left(\angle A\right)^{n} \;=\; A^{a}\Ab^{\,b},
\qquad a-b=n,\qquad a+b=m.
\label{eq:ab}
\end{equation}
Solving, $a = (m+n)/2$ and $b = (m-n)/2$. Now impose the Cardiff relation $m=|n|+2r$:
\begin{equation}
\boxed{\;\min(a,b) = r.\;}
\end{equation}
The identity is exact, not asymptotic. \emph{The index $r$ counts the conjugate pairs in
the term.} A term with $r=0$ is a pure power of $A$ or of $\Ab$; a term with $r=1$ carries
one factor of $|A|^2 = A\Ab$ alongside; and so on. The familiar restriction $r\le1$ retains
exactly the monomials with $\min(a,b)\le1$. Table~\ref{tab:dictionary} lists the low-order
terms in both notations.

\begin{table}[htbp]
\centering
\caption{The same terms in both notations, for $b_{2,1}$-type responses. The right-hand
columns are the ones that will acquire physical meaning in
Section~\ref{sec:corollaries}.}
\label{tab:dictionary}
\begin{tabular}{llrrrl}
\toprule
monomial & Cardiff term & $m$ & $n$ & $r$ & interpretation \\
\midrule
$1$              & constant          & 0 & 0  & 0 & no load-side dependence \\
$A$              & $|A|(\angle A)$   & 1 & 1  & 0 & linear in the load wave \\
$\Ab$            & $|A|(\angle A)^{-1}$ & 1 & $-1$ & 0 & linear, conjugate \\
$A^2$            & $|A|^2(\angle A)^2$  & 2 & 2  & 0 & second order \\
$A\Ab = |A|^2$   & $|A|^2$              & 2 & 0  & 1 & second order, one pair \\
$A^3$            & $|A|^3(\angle A)^3$  & 3 & 3  & 0 & third order \\
$A^2\Ab$         & $|A|^3(\angle A)$    & 3 & 1  & 1 & third order, one pair \\
$A^2\Ab^{\,2}$   & $|A|^4$              & 4 & 0  & 2 & \emph{excluded by $r\le1$} \\
\bottomrule
\end{tabular}
\end{table}

\subsection{The restriction $r\le1$ on the exponent lattice}

Figure~\ref{fig:lattice} plots the monomials on the $(a,b)$ lattice. Lines of constant $n$
run diagonally, lines of constant $m$ run anti-diagonally, and $r = \min(a,b)$ is the
distance from the nearer axis. The restriction $r\le1$ retains only the first two rows and the first two columns of
the lattice, an L-shaped region along the axes, and excludes the whole interior $a\ge2$,
$b\ge2$.

\begin{figure}[htbp]
\centering
\begin{tikzpicture}[scale=0.8]
  \fill[black!9] (-0.45,-0.45) rectangle (5.45,1.45);
  \fill[black!9] (-0.45,-0.45) rectangle (1.45,5.45);
  \draw[black!18,step=1] (-0.45,-0.45) grid (5.45,5.45);
  \draw[-{Stealth[length=2.5mm]}] (-0.45,0)--(5.95,0) node[right]{$a$};
  \draw[-{Stealth[length=2.5mm]}] (0,-0.45)--(0,5.95) node[above]{$b$};
  \foreach \i in {0,...,5}{
    \node[below,font=\scriptsize] at (\i,-0.5) {\i};
    \node[left,font=\scriptsize] at (-0.5,\i) {\i};}
  \draw[black!65,dashed,thick] (4.25,-0.25)--(-0.25,4.25);
  \draw[black!65,dotted,thick] (1.85,-0.15)--(5.35,3.35);
  \foreach \a in {0,...,5}{\foreach \b in {0,...,5}{
     \pgfmathtruncatemacro{\rr}{min(\a,\b)}
     \ifnum\rr<2
        \fill[black] (\a,\b) circle (2.3pt);
     \else
        \fill[white] (\a,\b) circle (2.6pt);
        \draw[black!60] (\a,\b) circle (2.6pt);
     \fi}}
  \draw[very thick] (2,2) circle (6.5pt);
  \begin{scope}[font=\footnotesize]
    \fill[black] (6.6,5.15) circle (2.3pt);
    \node[anchor=west] at (6.85,5.15) {kept by $r\le1$};
    \fill[white] (6.6,4.5) circle (2.6pt);
    \draw[black!60] (6.6,4.5) circle (2.6pt);
    \node[anchor=west] at (6.85,4.5) {discarded};
    \draw[black!65,dashed,thick] (6.35,3.5)--(6.95,3.5);
    \node[anchor=west] at (7.1,3.5) {$m=a+b=4$};
    \draw[black!65,dotted,thick] (6.35,2.85)--(6.95,2.85);
    \node[anchor=west] at (7.1,2.85) {$n=a-b=2$};
    \node[anchor=west,align=left] at (6.5,1.6)
       {$A^{2}\bar{A}^{2}=|A|^{4}$\\ first term discarded};
    \draw[black!70] (6.45,1.6)--(2.35,2.0);
  \end{scope}
\end{tikzpicture}
\caption{Monomials $A^{a}\Ab^{b}$ on the exponent lattice. The shaded L-shaped region is
$\min(a,b)\le1$, the Cardiff restriction $r\le1$: it contains every monomial through total
degree $3$, and $A^2\Ab^2=|A|^4$, the $|A|^4$ contribution to the $n=0$ coefficient, is
the first monomial it excludes.}
\label{fig:lattice}
\end{figure}

Counting is now simple. Through total degree $3$ the L-shaped region contains every
monomial, so the restriction excludes nothing. Beyond that the number of monomials retained by $r\le1$ grows linearly, $4D-2$, while
the number of all monomials of degree at most $D$ grows quadratically, $(D+1)(D+2)/2$:

\begin{center}
\begin{tabular}{rrrrr}
\toprule
degree $D$ & full dimension & kept by $r\le1$ & missing & fraction retained \\
\midrule
3  & 10 & 10 & 0  & 1.000 \\
4  & 15 & 14 & 1  & 0.933 \\
5  & 21 & 18 & 3  & 0.857 \\
9  & 55 & 34 & 21 & 0.618 \\
12 & 91 & 46 & 45 & 0.505 \\
\bottomrule
\end{tabular}
\end{center}

\noindent
The restriction therefore removes many coefficients at high order and almost none at low
order, which raises the question of when the excluded terms are identically zero.

\section{Bound on $r$}
\label{sec:proof}

\subsection{Binomial expansion of the load-side voltage}

Let the device be memoryless, with its dependence on the load-side voltage a polynomial
of degree $K$, and let the drive-side nonlinearity contribute harmonics of index $\ell$.
The memoryless hypothesis is used only here, because a polynomial degree in $v_2(t)$ is
a property of a static map. Let the load-side voltage be set by the incident
wave (open loop; Section~\ref{sec:loaded} removes this assumption), so that it is a
fundamental tone,
\begin{equation}
v_2(t) = \operatorname{Re}\left(Ae^{j\theta}\right),\qquad \theta=\omega_0 t .
\end{equation}
Raising it to the $p$-th power is an application of the binomial theorem:
\begin{equation}
v_2^{\,p} = 2^{-p}\sum_{q=0}^{p}\binom{p}{q} A^{q}\Ab^{\,p-q} e^{j(2q-p)\theta}.
\label{eq:binom}
\end{equation}
Read off the structure: a term of degree $p$ in the load-side voltage carries exactly $p$
factors drawn from $\{A,\Ab\}$ --- $q$ of one and $p-q$ of the other --- and it sits at
harmonic offset $2q-p$.

Extracting the $h$-th harmonic from the response requires that the drive-side contribution
$\ell$ and the load-side offset add up to $h$: $\ell + (2q-p) = h$. Solving alongside $a=q$, $b=p-q$:
\begin{equation}
a=\frac{p+h-\ell}{2},\qquad b=\frac{p-h+\ell}{2},
\qquad\Longrightarrow\qquad
\boxed{\,a+b=p,\qquad a-b=h-\ell.\,}
\label{eq:key}
\end{equation}

The boxed identities in \eqref{eq:key} are used in everything that follows. The first says the total
number of load-side factors equals the order of the load-side nonlinearity that produced
the term. The second says the imbalance between $A$'s and $\Ab$'s is set by the drive-side
harmonic.

\subsection{The conjugate-pair bound}

\begin{theorem}[Conjugate-pair bound, $r_{\max}=\lfloor K/2\rfloor$]\label{thm:rmax}
If a memoryless device's dependence on the load-side voltage is a polynomial of degree $K$, and
the load-side voltage is set by the incident wave, every term in the response satisfies
$\min(a,b)\le\lfloor K/2\rfloor$. Equivalently, in Cardiff indices,
\[
r_{\max}=\left\lfloor K/2 \right\rfloor .
\]
\end{theorem}

\begin{proof}
$\min(a,b)\ge r$ requires both $a\ge r$ and $b\ge r$. Adding and using $a+b=p$ from
\eqref{eq:key} gives $p\ge 2r$. Since $p\le K$ and $r$ is an integer, $r\le\lfloor
K/2\rfloor$.
\end{proof}

The proof is two lines and the statement is elementary. What it rests on is the reading
of $r$ as a count of conjugate pairs in the rewriting of~\cite{tian2025}
(Section~\ref{sec:monomials}); once $r$ is identified with $\min(a,b)$, the bound follows.
The theorem is a model-order rule for a device that satisfies its hypotheses: memoryless,
polynomial of degree $K$ in the load-side voltage, with that voltage set by the incident
wave. Section~\ref{sec:loaded} shows what replaces it when the last hypothesis fails.

\subsection{What Theorem~\ref{thm:rmax} says}

\begin{itemize}[leftmargin=1.4em]
\item \textbf{The restriction $r\le1$ holds for every term if and only if $K\le3$.} A
      cubic load-side nonlinearity cannot produce a term with two conjugate pairs, because
      two pairs need four factors and a cubic supplies at most three. So the Cardiff restriction is not an approximation for a cubic
      device --- it is exact. At fourth order it generically fails, and the first term to
      appear is $A^2\Ab^2 = |A|^4$, the circled point in Figure~\ref{fig:lattice}.
\item \textbf{The bound is attained, not merely respected.} A term with $\min(a,b)=r$ comes
      from $p=2r$ whenever the drive side supplies $\ell=h$, which any nonlinearity with a
      linear term does at the fundamental. So $\lfloor K/2 \rfloor$ is the actual maximum,
      not a loose upper bound.
\item \textbf{Separability is not assumed.} The argument uses only the degree in $v_2$, so
      a general non-separable $F(v_1,v_2)$ with cross terms obeys the same bound.
\item \textbf{Where the expansion is centered does not matter.} A-pull data is naturally
      expanded about the center of a circle rather than about $A=0$. Substituting $A\mapsto
      A+c$ into $A^{\alpha}\Ab^{\beta}$ produces only terms with $a\le\alpha$ and
      $b\le\beta$, so $\min(a,b)$ cannot increase. The bound is centering-independent.
      One feature does change under recentering, and Section~\ref{sec:apull} will need it: a
      source term of load-side order $p$, which about $A=0$ produces monomials with
      $a+b=p$ exactly, about $A=c$ produces monomials with $a+b\le p$. So in a centered
      expansion the degree of a term is a \emph{lower bound} on the order that fed it,
      while the highest degree present is still exactly $K$.
\end{itemize}

\begin{example}[A cubic device]
Take $K=3$ and ask for the fundamental response, $h=1$, with the drive side contributing
$\ell=1$. Then $a-b=0$, so $a=b$, and $a+b=p\le3$ forces $p\in\{0,2\}$ and
$(a,b)\in\{(0,0),(1,1)\}$. The response at the fundamental contains a constant and a
$|A|^2$ term: $r\le1$, as Theorem~\ref{thm:rmax} states. Raise $K$ to $4$ and
$p=4$ becomes available, admitting $(2,2)=|A|^4$ with $r=2$.
\end{example}

\subsection{Numerical check}

The result was checked for $K=2,\dots,7$ by brute-force expansion, both for separable and
non-separable $F$, using the companion MATLAB toolkit:

\begin{lstlisting}
>> T = xp_rmax_rule(2:7);      % expand, harmonic-select, tabulate min(a,b)
>> T = xp_device_demo(2:7);    % same sweep framed as "what does r<=1 cost?"
\end{lstlisting}

\noindent
Both report the observed $r_{\max}$ equal to $\lfloor K/2\rfloor$ at every order tested.
Appendix~\ref{app:matlab} lists the routines and what each one checks.

\section{What each exponent means}
\label{sec:corollaries}

The two identities \eqref{eq:key} give each of the three Cardiff indices a physical
meaning, stated as three corollaries.

\begin{corollary}[The magnitude exponent is the nonlinearity order]
$m=|n|+2r=a+b=p$. The magnitude exponent \emph{is} the order of the load-side
nonlinearity that produced the term.
\end{corollary}

\begin{corollary}[The phase exponent is set by the drive-side harmonic]
$n=a-b=h-\ell$. The phase exponent is set by which harmonic the drive-side nonlinearity
supplied.
\end{corollary}

\begin{corollary}[Mixing order]
$m+|n-h| = p+|\ell|$ is the lowest total nonlinearity order able to produce the term:
load-side order plus drive-side harmonic. This is the quantity tabulated as ``mixing
order'' in the A-pull identification literature~\cite{alrawachy2025}, and the identity can
be checked row by row against Table~2 of~\cite{alrawachy2025}.
\end{corollary}

\noindent
The physical interpretation of the three indices is summarized in
Table~\ref{tab:indices}.

\begin{table}[htbp]
\centering
\caption{Physical interpretation of the Cardiff indices, from the identities \eqref{eq:key}.}
\label{tab:indices}
\begin{tabular}{lll}
\toprule
index & equals & physical interpretation \\
\midrule
$m$ & $a+b=p$        & order of the load-side nonlinearity \\
$n$ & $a-b=h-\ell$   & harmonic supplied by the drive side \\
$r$ & $\min(a,b)$    & number of conjugate pairs; bounded by $\lfloor K/2\rfloor$ \\
$m+|n-h|$ & $p+|\ell|$ & mixing order \\
\bottomrule
\end{tabular}
\end{table}

\begin{translation}
The Cardiff literature limits $r$ to $0$ and $1$ ``because of extrapolation
concerns''~\cite{tian2025}. Theorem~\ref{thm:rmax} gives a second reading of the same
limit: it holds exactly when the device's load-side nonlinearity is at most cubic, a
hypothesis that an A-pull measurement with a low enough floor can test.
\end{translation}

\section{The loaded device: the decay in $r$}
\label{sec:loaded}

Theorem~\ref{thm:rmax} concerns the \emph{intrinsic} nonlinearity --- the device's own polynomial
dependence on its terminal voltage. In a real measurement the load-side voltage is not an
independent variable. It satisfies a self-consistency condition, $v_{ds}=a_2+b_2$, in which
$b_2$ on the right-hand side is the device's own output, so the load-side voltage is a
fixed point of that equation. The fixed point of a quadratic map is not a polynomial in
$a_2$ but a power series in it. So strictly speaking, a loaded device has no finite $K$, and $r_{\max}$ is infinite.

In place of a bound, the coefficients decrease with $r$: every value of $r$ is present, but each
unit increase in $r$ reduces the largest coefficient, by $1.1$ to $1.6$ decades at
$\rho=0.5$ and $1.7$ to $2.2$ decades at $\rho=0.25$ over the first five steps of the
example below. The rate of decrease depends on the amplitude of the load wave and on the
load. The series
is therefore truncated where the coefficients fall below the accuracy floor of the
measurement, rather than at a fixed $r$. Consider a device whose
intrinsic dependence on the load-side voltage is \emph{quadratic}, so that
Theorem~\ref{thm:rmax} gives $r_{\max}=1$:
\[
i_d(v_1,v_2) = 0.8\,v_1 + 0.2\,v_1^2 - 0.1\,v_1^3
             + 0.08\,v_1 v_2 + 0.04\,v_1^2 v_2 + 0.05\,v_1 v_2^2 + 0.04\,v_2^2 ,
\]
in $Z_0=1$ units with the drive $v_1=\cos\theta$ fixed. Inject only the fundamental at
port~2, $a_2(t)=\operatorname{Re}(Ae^{j\theta})$, and close the loop pointwise in time,
$b_2 = a_2 - i_d(v_1, a_2+b_2)$. Write $B_{2,1}$ as a series in the monomials
$A^{a}\Ab^{b}$, quote each coefficient in units of $\rho$ (so each is the size of that term
at the edge of the measured region $|A|\le\rho$), and record the largest coefficient in each
$r$-class relative to the constant term:

\begin{center}
\begin{tabular}{lcccccc}
\toprule
$r$ & 0 & 1 & 2 & 3 & 4 & 5 \\
\midrule
open loop, $\rho=0.5$ & $1$ & $8.6\times10^{-3}$ & $\sim10^{-14}$ & $\sim10^{-14}$
                      & $\sim10^{-14}$ & $\sim10^{-14}$ \\
closed loop, $\rho=0.5$ & $1$ & $3.8\times10^{-2}$ & $1.1\times10^{-3}$
                        & $5.8\times10^{-5}$ & $4.3\times10^{-6}$ & $3.7\times10^{-7}$ \\
closed loop, $\rho=0.25$ & $1$ & $9.4\times10^{-3}$ & $6.6\times10^{-5}$
                         & $9.1\times10^{-7}$ & $1.7\times10^{-8}$ & $3.6\times10^{-10}$ \\
\bottomrule
\end{tabular}
\end{center}

\noindent
The open-loop row is a least-squares fit over the disc and illustrates
Theorem~\ref{thm:rmax}: $K=2$, and everything beyond $r=1$ is zero to
machine precision. The closed-loop rows are exact. For this device the loop equation is a
quadratic in the load-side voltage at each instant, so its solution can be expanded in
closed form; the companion paper gives the formula~\cite{paper1}. A fit over the disc, which
is what \texttt{xp\_loaded\_demo} in the companion toolkit does and what one would do
with measured data, reproduces these numbers to within a few percent up to $r=4$ and
differs more at $r=5$.

The closed-loop rows show the effect of the loop: every value of $r$ is now present. Each
step in $r$ costs a factor $\rho^{2}$, from the extra conjugate pair $|A|^{2}$, times a
ratio of successive series coefficients that does not depend on $\rho$: $0.15$, $0.11$,
$0.22$, $0.29$ and $0.34$ for the first five steps. The ratio is not constant, so the
decay is not geometric over this range, and halving the load-wave amplitude divides each
step by four through $\rho^{2}$ alone.

In this example the load enters the decay through a single number. The series converges
for $|A|<R$, and $R$ depends on the load. Describing the same device in a reference impedance $Z$, the radius
of convergence is $R=4.10$, $2.15$, $1.29$, $0.81$ and $0.50$ for $Z=0.5$, $0.75$, $1$,
$1.25$ and $1.5$. From the third step on, the step ratios multiplied by $R^{2}$ are the
same for all five impedances to within a few percent, and as $r$ grows they rise toward
one ($0.96$ at the sixtieth step). The slope of the decay is therefore set by $\rho/R$, the size of the measured region
compared with the radius of convergence of the series, and the decay slows as the
load-pull radius approaches $R$.

The same example shows what the truncation costs. At $Z=1$ and $\rho=0.5$, keeping only
$r=0$ leaves a normalized mean-square error of $-34$\,dB over the disc, keeping $r\le1$
leaves $-67$\,dB, and each further value of $r$ lowers the error by $22$ to $26$\,dB. For
any device whose response is analytic there is a bound of the same form: the terms beyond
$r_{\max}$ sum to at most $M(\rho/R)^{2(r_{\max}+1)}/(1-\rho/R)^{2}$, where $M$ bounds the
response within the radius $R$~\cite{paper1}. For a measured device $M$ and $R$ are not
known in advance, so the bound tells the form of the truncation error and the fit tells
its size. The exact coefficients, the radius of convergence and the truncation errors are
computed by \texttt{loaded\_exact} in the companion toolkit.

The practical consequence is a change in the \emph{kind} of rule being applied. For an intrinsic nonlinearity,
$r_{\max}$ is set by a degree, specified by Theorem~\ref{thm:rmax}. For a loaded device,
$r_{\max}$ is set by where the decay crosses the accuracy floor, and the decay rate is
measurable from the fit itself.

\begin{figure}[htbp]
\centering
\begin{tikzpicture}
\begin{semilogyaxis}[
  width=11cm, height=6cm,
  xlabel={conjugate-pair index $r$},
  ylabel={largest coefficient (relative)},
  xmin=-0.3, xmax=4.3, ymin=3e-7, ymax=4,
  xtick={0,1,2,3,4},
  grid=both, grid style={black!12},
  tick label style={font=\footnotesize},
  label style={font=\small},
  legend style={font=\footnotesize, draw=black!30, at={(0.98,0.98)}, anchor=north east}]
\addplot[black, thick, mark=*, mark size=2pt]
  coordinates {(0,1) (1,3.8e-2) (2,1.1e-3) (3,5.8e-5) (4,4.3e-6)};
\addlegendentry{closed loop, $\rho=0.5$}
\addplot[black!45, thick, mark=o, mark size=2pt]
  coordinates {(0,1) (1,9.4e-3) (2,6.6e-5) (3,9.1e-7)};
\addlegendentry{closed loop, $\rho=0.25$}
\addplot[black!55, dashed, thick, domain=-0.3:4.3, samples=2]
  {1e-4};
\addlegendentry{illustrative floor, $10^{-4}$}
\node[font=\footnotesize, anchor=west] at (axis cs:3.2,1.6e-3) {truncate here};
\draw[-{Stealth[length=2mm]}, black!70] (axis cs:3.35,9.0e-4) -- (axis cs:3.06,1.5e-4);
\end{semilogyaxis}
\end{tikzpicture}
\caption{The stopping rule for a loaded device (exact coefficients of the example).
Plot the largest coefficient at each $r$ on a log axis and truncate where the decay
meets the floor. The slope is set by the ratio of the load-wave amplitude to the radius of
convergence of the series. The floor drawn at $10^{-4}$ is illustrative.}
\label{fig:decay}
\end{figure}

\noindent
The rule is: fit, plot coefficient magnitude against $r$ on a log axis, and
truncate where it meets the noise (Figure~\ref{fig:decay}). The argument has the same form as the truncation of a
Volterra series or an FIR tail: the order is set by the floor, not by a theorem. Theorem~\ref{thm:rmax} adds the intrinsic behavior: it states what the device would produce
without the loop, so a decay that stops at low $r$ says that the device's load-side
nonlinearity is of low degree, not that the fit was truncated there.

\section{Measuring the decomposition}
\label{sec:apull}

\subsection{Why an ordinary load-pull sweep does not show the decomposition}

In a conventional load-pull sweep every term in \eqref{eq:cardiff} contributes to every
measured point. The coefficients are recovered by least squares, so the answer depends on
the basis assumed, and when two basis functions are nearly collinear over the measured
load states the fit can assign the response of one exponent pair to another. Individual $(m,n)$ terms are not detectable, only
their sum.

\subsection{The tailored A-pull method}

The tailored A-pull method of Tasker and co-workers~\cite{alrawachy2025,tasker2020}
separates the terms by modulating the load wave. Instead of visiting load states in an
arbitrary order, the load-side wave is stepped around a \emph{spiral} about a chosen center $a_{2,1}^{\mathrm{c}}$:
\[
a_{2,1}(I) = a_{2,1}^{\mathrm{c}} + \tfrac{1}{2}\Delta a\left[1+\cos\!\left(\tfrac{2\pi
S_{am} I}{N}\right)\right] e^{\,j 2\pi S_{pm} I/N},\qquad I=0,\dots,N-1,
\]
which is Eq.~(1) of~\cite{alrawachy2025}: the deviation from the center is amplitude modulated at rate
$S_{am}$ (a raised cosine, so the magnitude sweeps from $0$ to $\Delta a$) and phase
modulated at rate $S_{pm}$, both as functions of the sequence index $I$. The measured
response $b_{2,1}(I)$ is then Fourier transformed over the sequence index. The model of~\cite{alrawachy2025}, its Eq.~(7), is written in the deviation $A \equiv
a_{2,1}-a_{2,1}^{\mathrm{c}}$, which is the centered expansion of
Section~\ref{sec:proof}.

The separation works as follows.
Consider a single term $A^{a}\Ab^{\,b} = |A|^{m}e^{jn\phi}$ in that centered variable and
follow it around the spiral.

\begin{itemize}[leftmargin=1.4em]
\item The \textbf{phase} part contributes $e^{jn\phi}$, and $\phi$ advances linearly at
      rate $S_{pm}$. That is a single spectral line at $n\,S_{pm}$. The position of the
      line \emph{is} the exponent $n$.
\item The \textbf{magnitude} part contributes $|A|^{m}$, and $|A|$ is a raised cosine
      at rate $S_{am}$. Raising $[1+\cos x]$ to the $m$-th power generates harmonics of
      $x$ up to order $m$ and no further --- the same bookkeeping as \eqref{eq:binom},
      applied to the modulation instead of the carrier. So the line at $n\,S_{pm}$
      acquires sidebands at $\pm s\,S_{am}$ for $s\le m$.
\end{itemize}

The result is a \emph{cluster} of lines at $s\,S_{am} + n\,S_{pm}$, $s=-m,\dots,m$
--- Eq.~(8) of~\cite{alrawachy2025} --- centered at $n\,S_{pm}$ with half-width $m\,S_{am}$. The center of the cluster gives $n$ and its half-width gives $m$. The spectrum is a direct picture of the exponent
decomposition of Section~\ref{sec:monomials}.
Figure~\ref{fig:spectrum} is the schematic. The two rates $S_{am}$ and $S_{pm}$ are chosen so that clusters of
neighboring $n$ do not overlap: since a cluster at $n\,S_{pm}$ can be as wide as $K\,S_{am}$
on either side, Al-Rawachy \emph{et al.}~\cite{alrawachy2025} require $S_{pm}>(2\,\mathrm{Order}-1)\,S_{am}$, their
Eq.~(9), where Order is the largest mixing order sought, and take $S_{am}$ and $S_{pm}$
prime, so that no line of one cluster falls on a line of another; their values are
$S_{am}=3$ or $5$ and $S_{pm}=53$ or $89$.

\begin{figure}[htbp]
\centering
\begin{tikzpicture}[yscale=0.055, xscale=1.0]
  \draw[-{Stealth[length=2.5mm]}] (-0.4,0)--(11.2,0) node[below right,font=\footnotesize]{frequency};
  \draw[-{Stealth[length=2.5mm]}] (-0.4,0)--(-0.4,68);
  \node[font=\footnotesize,rotate=90,anchor=south] at (-0.95,34) {line amplitude};
  \foreach \i/\hh in {-2/14,-1/30,0/60,1/30,2/14}{
     \draw[very thick] ({2.0+0.28*\i},0)--({2.0+0.28*\i},\hh);}
  \draw[decorate,decoration={brace,amplitude=4pt,mirror},black!70]
     (1.44,-6)--(2.56,-6);
  \node[font=\footnotesize,align=center,anchor=north] at (2.0,-12) {$(m,n)=(2,0)$\\ half-width $2$};
  \foreach \i/\hh in {-3/7,-2/13,-1/24,0/42,1/24,2/13,3/7}{
     \draw[very thick] ({6.0+0.28*\i},0)--({6.0+0.28*\i},\hh);}
  \draw[decorate,decoration={brace,amplitude=4pt,mirror},black!70]
     (5.16,-6)--(6.84,-6);
  \node[font=\footnotesize,align=center,anchor=north] at (6.0,-12) {$(3,1)$\\ half-width $3$};
  \foreach \i/\hh in {-2/6,-1/12,0/22,1/12,2/6}{
     \draw[very thick] ({9.4+0.28*\i},0)--({9.4+0.28*\i},\hh);}
  \draw[decorate,decoration={brace,amplitude=4pt,mirror},black!70]
     (8.84,-6)--(9.96,-6);
  \node[font=\footnotesize,align=center,anchor=north] at (9.4,-12) {$(2,2)$};
  \draw[{Stealth[length=2mm]}-{Stealth[length=2mm]},black!70] (2.0,66)--(6.0,66);
  \node[font=\footnotesize,anchor=south] at (4.0,66) {spacing $=S_{pm}$};
  \draw[{Stealth[length=2mm]}-{Stealth[length=2mm]},black!70] (5.72,50)--(6.0,50);
  \node[font=\footnotesize,anchor=west] at (6.1,50) {line spacing $=S_{am}$};
\end{tikzpicture}
\caption{Schematic of a tailored A-pull spectrum. Each Cardiff term appears as a cluster:
its center frequency encodes the phase exponent $n$, its half-width encodes the magnitude
exponent $m$. By Corollary~1 and the centering remark of Section~\ref{sec:proof}, the
widest cluster present has half-width $K$ --- the spectrum gives the order of the
load-side nonlinearity.}
\label{fig:spectrum}
\end{figure}

\subsection{The cluster width has a physical interpretation}

Corollary~1 said $m=p$ for an expansion about zero load wave, and the centering remark
in Section~\ref{sec:proof} said that about any other center a source of order $p$ feeds
degrees $m\le p$, with the top degree $K$ still reached. Combining:

\begin{center}
\emph{The half-width of a cluster is the lowest load-side order able to produce it, and the
widest cluster in the spectrum has half-width equal to the order $K$ of the load-side
nonlinearity.}
\end{center}

\noindent
This statement converts a spectral feature into a device parameter. The half-width of the widest resolvable cluster gives $K$, and Theorem~\ref{thm:rmax}
then gives $r_{\max}=\lfloor K/2\rfloor$.

\subsection{What the published data shows}

Al-Rawachy \emph{et al.}~\cite{alrawachy2025} applied the tailored A-pull identification
to a GaN device at 2.45\,GHz and list the extracted coefficients for $b_{2,1}$ in their Table~2. The
first four columns of Table~\ref{tab:alrawachy} below are theirs; the last two are what
this tutorial adds.

\begin{table}[htbp]
\centering
\caption{The nine exponent pairs identified for $b_{2,1}$ on a GaN device at 2.45\,GHz.
Columns $m$, $n$, mixing order and coefficient magnitude are from Table~2
of~\cite{alrawachy2025} (the $S_{am}=5$, $S_{pm}=89$, $N=1069$ column). The last two are
computed here: $r=(m-|n|)/2$ from Section~\ref{sec:monomials}, and the mixing order
of Corollary~3, $m+|n-h|$ with $h=1$.}
\label{tab:alrawachy}
\begin{tabular}{rrcrrc}
\toprule
$m$ & $n$ & mixing order (theirs) & $|C_{2,1,m,n}|$ & $r=\frac{m-|n|}{2}$
    & $m+|n-h|$ (Cor.~3) \\
\midrule
0 & $0$  & 1 & 1.379 & 0 & 1 \\
1 & $1$  & 1 & 0.462 & 0 & 1 \\
1 & $-1$ & 3 & 0.075 & 0 & 3 \\
2 & $0$  & 3 & 0.050 & 1 & 3 \\
2 & $2$  & 3 & 0.039 & 0 & 3 \\
2 & $-2$ & 5 & 0.007 & 0 & 5 \\
3 & $1$  & 3 & 0.027 & 1 & 3 \\
3 & $-1$ & 5 & 0.015 & 1 & 5 \\
3 & $3$  & 5 & 0.008 & 0 & 5 \\
\bottomrule
\end{tabular}
\end{table}

Three points follow from the two computed columns.

\begin{itemize}[leftmargin=1.4em]
\item \textbf{Every pair has $r\le1$}, and the largest magnitude exponent is $m=3$. No
      pair with $m\ge4$ appears --- in particular neither $(4,0)$ nor $(5,1)$, which are
      the two $r=2$ pairs that Al-Rawachy \emph{et al.}~\cite{alrawachy2025} report
      searching for and not finding.
\item \textbf{The table does not test Theorem~\ref{thm:rmax}.} By Corollary~1 the
      load-side order detected above the floor of that measurement is $3$, and
      Theorem~\ref{thm:rmax} then gives $r_{\max}=\lfloor3/2\rfloor=1$, which is what the
      table shows. But $K$ has been read from the largest $m$ in the same table, and every
      table whose largest $m$ is $3$ has $r\le1$. A test needs $K$ to be known
      independently, as it is in the simulations below, or a measurement whose floor is
      lowered until the first term with $m\ge4$ appears.
\item \textbf{The mixing-order identity checks on every row.} The last column, computed
      from Corollary~3, reproduces the tabulated mixing
      order in all nine rows, which checks that the two definitions of mixing order agree.
\end{itemize}

\subsection{Extrapolating to the $r=2$ term}

Table~\ref{tab:alrawachy} also gives a rough indication of where the first $r=2$ term may
lie. A caution on units comes first. The coefficients of~\cite{alrawachy2025} multiply
$|A|^{m}$ with $A$ in the wave units of that paper, and the levels quoted here take
$|A|=1$ in those units; at another amplitude a term of order $m$ scales as $|A|^{m}$.
Comparing the $r=1$ coefficient with the $r=0$ coefficient at the same $n$:
\[
n=0:\;\frac{0.050}{1.379}=0.036,\qquad
n=1:\;\frac{0.027}{0.462}=0.058,\qquad
n=-1:\;\frac{0.015}{0.075}=0.20 .
\]
Now extrapolate one step further. If the step from $(2,0)$ to $(4,0)$, the first $r=2$
term, had one of these ratios, the $(4,0)$ coefficient would be $0.050$ times that ratio,
and referred to the $(0,0)$ coefficient of $1.379$ it would lie at
\[
-58,\qquad -54\qquad\text{or}\qquad -43\ \text{dBc}.
\]
The three figures differ by $15$\,dB, and a one-step extrapolation of this kind is not
reliable in any case: in the simulation of Section~\ref{sec:asmhemt} it is low by $6$ to
$16$\,dB.

The identification floor of~\cite{alrawachy2025} is a signal-to-spurious ratio near $-40$\,dBc with a noise floor
near $-60$\,dBc. The estimate is therefore a range, from about $-58$\,dBc up to the
spurious floor, and the fact that $(4,0)$ was not detected is consistent with every level
in that range. Al-Rawachy \emph{et al.} draw the qualitative conclusion themselves: the
dynamic range of the measurement was not sufficient to extract all of the fifth-order
terms accurately~\cite{alrawachy2025}.

The closed-loop calculation of Section~\ref{sec:loaded} shows the mechanism by which such
a term can arise in a device whose intrinsic nonlinearity is of lower order: fourth-order
content generated through the load. Its level there, $1.1\times10^{-3}$ relative to the
constant at $\rho=0.5$, moves as the fourth power of the load-wave amplitude and with the
load, and no significance attaches to its closeness to the low end of the range above.

\subsection{Resolving the $r=2$ term}

If the spurious floor can be pushed toward the
$-60$\,dBc noise floor --- better phase-reference stability, more cycles, spurious-line
cancellation --- then:

\begin{itemize}[leftmargin=1.4em]
\item the $(4,0)$ cluster should be the first new term to appear, somewhere in the range
      above, with $(5,1)$ the next $r=2$ term behind it. The level is uncertain; the
      \emph{ordering} is structural: any $r=2$ term needs
      load-side order at least four, and $(4,0)$ is the only one that needs no more than
      four. The fourth-order $r=1$ term $(4,2)$, also absent from Table~2 of~\cite{alrawachy2025}, may
      appear in between; the simulation below puts it there;
\item if instead nothing appears down to $-60$\,dBc, then $r\le1$ is confirmed for that
      device at that floor by direct measurement rather than imposed, and the effective $K$ is
      $\le3$ at that drive and mismatch.
\end{itemize}

\noindent
Either outcome is informative, and the measurement can be made with the apparatus
described in Section~III of~\cite{alrawachy2025}.

\subsection{A simulated run-through}
\label{sec:apullsim}

Everything in this section can be simulated on a synthetic device with the routine \texttt{xp\_apull\_demo} in the companion toolkit. The device
is a Curtice-type transistor in $Z_0=1$ units, $i_d = g\,G(v_1)\,F_K(v_2)$ with
$G(v_1)=\tanh(1.5\,(0.2+v_1))$ on the drive side and $F_K$ the degree-$K$ Taylor
polynomial of $\tanh(1+v_2)$ on the load side, so that $K$ is the intrinsic load-side
order and Theorem~\ref{thm:rmax} gives $r_{\max}=\lfloor K/2\rfloor$. The device is loaded, that is, the load-side voltage is the fixed point of
$v_{ds}=a_2+b_2$ as in Section~\ref{sec:loaded}, so that fourth-order content can arise
through the load. The stimulus is Eq.~(1) of~\cite{alrawachy2025} with the published
numbers, $S_{am}=5$, $S_{pm}=89$, six cycles, $N=1069$ points, load-wave amplitude up
to half the drive; the response $b_{2,1}$ is Fourier transformed over the sequence
index, and a pair $(m,n)$ is called detected when its outermost line
$n\,S_{pm}\pm m\,S_{am}$ stands $3$\,dB above a chosen floor.

\begin{figure}[htbp]
\centering
\begin{tikzpicture}
\begin{axis}[
  width=15cm, height=7cm,
  xlabel={bin index over the A-pull sequence, $n\,S_{pm}+s\,S_{am}$},
  ylabel={level re $(0,0)$, dBc},
  xmin=-230, xmax=230, ymin=0, ymax=96,
  ytick={0,20,40,60,80}, yticklabels={$-90$,$-70$,$-50$,$-30$,$-10$},
  xtick={-178,-89,0,89,178}, xticklabels={$-2S_{pm}$,$-S_{pm}$,$0$,$S_{pm}$,$2S_{pm}$},
  grid=major, grid style={black!12},
  tick label style={font=\footnotesize}, label style={font=\small},
  clip=true]
\addplot[ycomb, black, line width=0.6pt, mark=none] coordinates {
(-227,1.3) (-223,1.8) (-218,5.0) (-213,8.5) (-208,12.5) (-203,16.9) (-198,21.8) (-193,27.1) (-188,22.8) (-183,33.2) (-178,39.0) (-173,33.2) (-168,22.8) (-163,27.1) (-158,21.8) (-153,16.8) (-148,12.4) (-143,8.2) (-139,0.2) (-138,4.5) (-134,3.4) (-133,1.1) (-129,7.0) (-124,11.1) (-119,15.8) (-114,21.2) (-109,27.4) (-104,35.9) (-99,44.8) (-94,49.3) (-89,50.4) (-84,49.3) (-79,44.8) (-74,35.9) (-69,27.4) (-64,21.1) (-59,15.7) (-54,11.0) (-50,0.8) (-49,6.8) (-45,4.3) (-44,3.0) (-40,8.2) (-35,12.6) (-30,17.6) (-25,23.4) (-20,30.4) (-15,38.2) (-10,53.5) (-5,63.4) (0,90.0) (5,63.4) (10,53.5) (15,38.2) (20,30.4) (25,23.4) (30,17.6) (35,12.5) (39,0.9) (40,8.1) (44,4.4) (45,4.2) (49,8.4) (50,0.6) (54,12.9) (59,17.9) (64,23.9) (69,30.8) (74,40.7) (79,50.6) (84,78.4) (89,84.2) (94,78.4) (99,50.6) (104,40.7) (109,30.8) (114,23.9) (119,18.0) (124,12.9) (128,0.3) (129,8.4) (133,3.8) (134,4.5) (138,7.6) (139,0.9) (143,11.9) (148,16.8) (153,22.3) (158,28.9) (163,36.2) (168,49.0) (173,58.1) (178,61.1) (183,58.1) (188,49.0) (193,36.2) (198,28.9) (203,22.3) (208,16.8) (213,12.0) (218,7.7) (222,2.3) (223,3.9) (227,5.9) (228,0.5)
};
\addplot[black!60, dashed, thick] coordinates {(-230,50) (230,50)};
\addplot[black!60, dotted, thick] coordinates {(-230,30) (230,30)};
\node[font=\scriptsize, anchor=south west, fill=white, inner sep=1pt] at (axis cs:-228,50) {spurious floor, $-40$\,dBc};
\node[font=\scriptsize, anchor=south west, fill=white, inner sep=1pt] at (axis cs:-228,30) {noise floor, $-60$\,dBc};
\node[font=\scriptsize, anchor=south, fill=white, inner sep=1pt] at (axis cs:14,88)   {$(0,0)$};
\node[font=\scriptsize, anchor=south, fill=white, inner sep=1pt] at (axis cs:89,79)   {$(1,1)$};
\node[font=\scriptsize, anchor=south, fill=white, inner sep=1pt] at (axis cs:-89,57)  {$(1,-1)$};
\node[font=\scriptsize, anchor=south, fill=white, inner sep=1pt] at (axis cs:178,57)  {$(2,2)$};
\node[font=\scriptsize, anchor=south, fill=white, inner sep=1pt] at (axis cs:30,55)   {$(2,0)$};
\node[font=\scriptsize, anchor=south, fill=white, inner sep=1pt] at (axis cs:128,45)  {$(3,1)$};
\node[font=\scriptsize, anchor=south, fill=white, inner sep=1pt] at (axis cs:-60,37)  {$(3,-1)$};
\node[font=\scriptsize, anchor=south, fill=white, inner sep=1pt] at (axis cs:-40,26)  {$(4,0)$};
\end{axis}
\end{tikzpicture}
\caption{Simulated tailored A-pull spectrum for the loaded cubic ($K=3$) device,
$|k|\le230$ shown. Each cluster is centered at $n\,S_{pm}$ and its half-width
in units of $S_{am}=5$ is $m$: the $(2,0)$ cluster spans $\pm10$ bins about zero, the
$(3,1)$ cluster $\pm15$ about $89$. The two floors of~\cite{alrawachy2025} are drawn
in. The $(4,0)$ line at $\pm20$ bins, $-59.6$\,dBc, is the first fourth-order content and
sits at the noise floor.}
\label{fig:apullsim}
\end{figure}

Figure~\ref{fig:apullsim} is the spectrum, which can be compared with Fig.~6
of~\cite{alrawachy2025}.
Sweeping the identification floor and recording what is detected gives:

\begin{center}
\begin{tabular}{lp{0.34\textwidth}p{0.40\textwidth}}
\toprule
floor & newly detected pairs & interpretation \\
\midrule
$-40$\,dBc & $(0,0)$, $(1,1)$, $(2,0)$ & \\
$-50$\,dBc & $+\,(1,-1)$, $(2,2)$ & \\
$-60$\,dBc & $+\,(3,1)$, $(3,-1)$ & seven pairs, all $r\le1$, max $m=3$: $K_{\rm eff}=3$,
             $r_{\max}=1$ \\
$-70$\,dBc & $+\,(3,3)$, $(4,0)$, $(4,2)$, $(5,1)$, $(2,-2)$ & $(4,0)$ first, at
             $-59.6$\,dBc; then $(4,2)$ at $-61.1$\,dBc; then $(5,1)$ at $-66.1$\,dBc \\
\bottomrule
\end{tabular}
\end{center}

\noindent
At a $-60$\,dBc floor the synthetic device gives seven of
the nine pairs in Table~\ref{tab:alrawachy}, every one with $r\le1$, largest magnitude
exponent three, and so the inference $K_{\rm eff}=3$, $r_{\max}=1$. Lowering the floor by
another ten decibels brings in the fourth-order content in the order argued above ---
$(4,0)$ leading, then the $r=1$ term $(4,2)$, then $(5,1)$. The polynomial device puts $(4,0)$ at
$-59.6$\,dBc. That level moves with the load-wave amplitude and with the load, and no
significance attaches to its closeness to the lower end of the range estimated from the
measured coefficients. The structure does not move: the same run with the
loop opened gives $(4,0)$ at $-300$\,dBc for $K=3$ (a numerical zero, as Theorem~\ref{thm:rmax} requires)
and at $-109$\,dBc for $K=4$. In this device, essentially all of the fourth-order content
at the fundamental is generated through the load, which is the mechanism
Section~\ref{sec:loaded} described.

\texttt{xp\_apull\_demo} takes as input $K$, the open/closed switch, the floor and the
stimulus parameters, so the experiment of the previous subsection can be simulated for any
polynomial device before it is run on a real one.

\subsection{A compact transistor model in a circuit simulator}
\label{sec:asmhemt}

The polynomial device of the previous subsection has the right structure but is not a
realistic transistor. The same experiment was therefore repeated on a physics-based compact model in an
open-source circuit simulator: the ASM-HEMT model of a GaN high-electron-mobility
transistor (HEMT) (version 101.4.0 of the model
described in~\cite{khandelwal2018asmhemt}; Verilog-A),
compiled with OpenVAF and run in ngspice~44 through its OSDI interface. The default
parameter set describes a 0.8\,mm device with $V_{\rm off}=-2$\,V; it was biased at
$V_{gs}=-1.5$\,V (class AB, $\approx170$\,mA) with about 20\,V on the drain, driven at
2.45\,GHz from a 50\,$\Omega$ source, and loaded by a 50\,$\Omega$ environment. Self-heating
and trapping were switched off so that the device is memoryless, as Theorem~\ref{thm:rmax}
assumes, and reaches its CW steady state within a few periods.
A-pull is simple in simulation, as Al-Rawachy
\emph{et al.} note: a Thevenin source $2\sqrt{Z_0}\operatorname{Re}(A_{2}e^{j\omega t})$
behind $Z_0$ at the drain sets the incident wave $a_{2}$ exactly, whatever the device does.
The maximum-power load was found by a scan, the spiral of Eq.~(1) was wound about it with
$S_{am}=5$, $S_{pm}=89$, six cycles and $N=1069$ points, and its radius was chosen to
enclose roughly the 3\,dB output-power contour, the criterion of~\cite{alrawachy2025}. Each
point is a short transient (eight periods, the last four analyzed) at a fixed step of
$1024$ per period. The scripts, the
netlist and both datasets are in the \texttt{spice/} folder of the companion
toolkit~\cite{toolkit}, and
\texttt{xp\_apull\_asmhemt} reads the data into MATLAB and hands it to
\texttt{xp\_apull\_read}, the analysis half of \texttt{xp\_apull\_demo}.

\begin{figure}[htbp]
\centering
\begin{tikzpicture}
\begin{axis}[
  width=15cm, height=7cm,
  xlabel={bin index over the A-pull sequence, $n\,S_{pm}+s\,S_{am}$},
  ylabel={level re $(0,0)$, dBc},
  xmin=-230, xmax=230, ymin=0, ymax=106,
  ytick={0,20,40,60,80,100}, yticklabels={$-100$,$-80$,$-60$,$-40$,$-20$,$0$},
  xtick={-178,-89,0,89,178}, xticklabels={$-2S_{pm}$,$-S_{pm}$,$0$,$S_{pm}$,$2S_{pm}$},
  grid=major, grid style={black!12},
  tick label style={font=\footnotesize}, label style={font=\small},
  clip=true]
\addplot[ycomb, black, line width=0.6pt, mark=none] coordinates {
(-227,0.9) (-218,2.1) (-208,13.0) (-203,5.1) (-198,25.1) (-194,6.9) (-189,13.4) (-188,24.6) (-184,16.3) (-183,48.8) (-179,17.2) (-178,53.8) (-174,16.3) (-173,48.8) (-169,13.4) (-168,24.6) (-164,6.8) (-158,25.1) (-153,5.1) (-148,12.9) (-138,2.2) (-129,2.9) (-119,14.0) (-114,7.9) (-109,28.8) (-104,17.4) (-100,8.2) (-99,52.2) (-95,12.7) (-94,68.0) (-93,5.5) (-90,14.1) (-89,75.3) (-88,7.2) (-85,12.8) (-84,68.0) (-83,5.5) (-80,8.2) (-79,52.2) (-74,17.4) (-69,28.8) (-64,7.9) (-59,14.0) (-49,2.9) (-40,3.1) (-30,14.3) (-25,9.0) (-20,28.2) (-15,34.6) (-10,55.9) (-9,8.2) (-6,1.6) (-5,68.4) (-4,12.2) (-1,4.2) (0,100.0) (1,13.4) (4,1.7) (5,68.4) (6,12.2) (10,55.9) (11,8.2) (15,34.6) (20,28.2) (25,9.0) (30,14.3) (40,3.2) (49,2.8) (59,13.7) (64,11.6) (69,27.7) (74,31.8) (75,4.6) (78,1.1) (79,41.5) (80,10.5) (83,0.5) (84,77.4) (85,13.1) (89,83.3) (90,13.8) (93,0.4) (94,77.4) (95,13.1) (98,1.0) (99,41.5) (100,10.5) (103,0.1) (104,31.8) (105,4.6) (109,27.7) (114,11.5) (119,13.7) (129,2.8) (138,2.0) (148,12.2) (153,12.4) (158,24.5) (163,33.3) (164,5.0) (167,2.7) (168,49.2) (169,5.2) (172,4.7) (173,62.4) (177,5.3) (178,66.0) (182,4.7) (183,62.4) (187,2.6) (188,49.2) (189,5.1) (193,33.3) (194,5.0) (198,24.5) (203,12.4) (208,12.2) (218,2.0) (227,0.3)
};
\addplot[black!60, dashed, thick] coordinates {(-230,60) (230,60)};
\addplot[black!60, dotted, thick] coordinates {(-230,40) (230,40)};
\node[font=\scriptsize, anchor=south west, fill=white, inner sep=1pt] at (axis cs:-228,60) {spurious floor, $-40$\,dBc};
\node[font=\scriptsize, anchor=south west, fill=white, inner sep=1pt] at (axis cs:-228,40) {noise floor, $-60$\,dBc};
\node[font=\scriptsize, anchor=south, fill=white, inner sep=1pt] at (axis cs:14,99)   {$(0,0)$};
\node[font=\scriptsize, anchor=south, fill=white, inner sep=1pt] at (axis cs:89,80)   {$(1,1)$};
\node[font=\scriptsize, anchor=south, fill=white, inner sep=1pt] at (axis cs:-89,71)  {$(1,-1)$};
\node[font=\scriptsize, anchor=south, fill=white, inner sep=1pt] at (axis cs:30,58)   {$(2,0)$};
\node[font=\scriptsize, anchor=south, fill=white, inner sep=1pt] at (axis cs:178,52)  {$(2,2)$};
\node[font=\scriptsize, anchor=south, fill=white, inner sep=1pt] at (axis cs:125,35)  {$(3,1)$};
\node[font=\scriptsize, anchor=south, fill=white, inner sep=1pt] at (axis cs:-40,31)  {$(4,0)$};
\end{axis}
\end{tikzpicture}
\caption{Tailored A-pull spectrum of the ASM-HEMT GaN model in ngspice, $b_{2,1}$ at the
softer of the two drives (about 1--2\,dB into compression), $|k|\le230$ shown, with the two
floors of~\cite{alrawachy2025} drawn in. Lines that sit on no cluster position with
$|n|\le8$ are below $-80$\,dBc. Compare Figures~\ref{fig:apullsim}
and~\ref{fig:spectrum}.}
\label{fig:asmhemt}
\end{figure}

Figure~\ref{fig:asmhemt} is the spectrum at the softer drive, and Table~\ref{tab:asmhemt}
collects the outermost-line levels at both drives, arranged by $n$ so that the decay in $r$
can be read along each row. The levels come from a single simulator and a single method,
a transient run to steady state at each point; they have not been cross-checked by
harmonic balance or against a measurement.

\begin{table}[htbp]
\centering
\caption{ASM-HEMT: outermost-line level of each $(m,n)$ cluster in dBc relative to
$(0,0)$, at two drive levels. Along a row $m$ increases by two, i.e.\ $r$ by one. Entries
in parentheses lie below the largest line that sits on no cluster position with
$|n|\le8$, $-80$\,dBc at the softer drive and $-107$\,dBc at the harder, and are not to be
read as device response. Counting clusters to $|n|\le6$ only, as \texttt{xp\_apull\_read}
does, that level is $-72$\,dBc and $-92$\,dBc.}
\label{tab:asmhemt}
\begin{tabular}{lrrrcrrr}
\toprule
& \multicolumn{3}{c}{drive 0.8\,V (1--2\,dB compressed)} & &
  \multicolumn{3}{c}{drive 1.2\,V (3--4\,dB compressed)} \\
\cmidrule{2-4}\cmidrule{6-8}
$n$ & $m=|n|$ & $|n|{+}2$ & $|n|{+}4$ & & $m=|n|$ & $|n|{+}2$ & $|n|{+}4$ \\
\midrule
$0$  & $0.0$   & $-44.1$ & $-71.8$   & & $0.0$   & $-51.8$ & $-97.6$  \\
$+1$ & $-22.6$ & $-68.2$ & $(-88.4)$ & & $-27.2$ & $-81.1$ & $(-116)$ \\
$-1$ & $-32.0$ & $(-82.6)$ & $(-92.1)$ & & $-30.8$ & $-71.8$ & $(-113)$ \\
$+2$ & $-50.8$ & $-75.5$ & $(-87.8)$ & & $-57.2$ & $-100$  & $(-131)$ \\
$-2$ & $-75.4$ & $-74.9$ & $(-87.0)$ & & $-56.8$ & $-91.3$ & $(-131)$ \\
$+3$ & $-74.5$ & $(-86.9)$ & $(-101)$ & & $-80.3$ & $(-118)$ & $(-150)$ \\
$-3$ & $-73.7$ & $(-92.4)$ & $(-107)$ & & $-87.3$ & $(-110)$ & $(-141)$ \\
\bottomrule
\end{tabular}
\end{table}

Four points can be made about the ASM-HEMT spectra.

\begin{itemize}[leftmargin=1.4em]
\item \textbf{The lines above the floor sit on cluster positions.} Counting clusters to
      $|n|\le6$, the positions $n\,S_{pm}+s\,S_{am}$ cover $311$ of the $1069$ bins, and
      the largest line off that grid is at $-72$\,dBc at the softer drive and $-92$\,dBc
      at the harder. Those lines sit on the positions of the $|n|=7$ and $8$ clusters,
      which fall one bin from the centers of the $n=\mp5$ and $\mp4$ clusters because
      $12\,S_{pm}=N-1$. Counting clusters to $|n|\le8$ ($471$ bins), the largest remaining
      line is at $-80$\,dBc and $-107$\,dBc. Since $S_{am}$ and $N$ are coprime, the
      cluster positions fill every bin as the order is raised, so this check has force
      only at low order. At the softer drive the center lines of the clusters with $n=4$
      to $7$ lie between $-64$ and $-72$\,dBc and do not fall with $n$, whereas at the
      harder drive they fall by about $10$\,dB per unit of $n$. Rerunning the softer
      drive with twice the steps per period, a tenfold smaller relative tolerance and
      twelve periods in place of eight changes no spectral line by more than $-107$\,dB
      relative to $(0,0)$, and the softer-drive levels therefore come from the model as
      simulated. A likely cause is that the softer-drive spiral straddles the onset of knee
      clipping. The minimum drain voltage in a period ranges from $0.05$ to $9$\,V across
      that spiral, whereas at the harder drive it is below $0.75$\,V at every point
      checked. This explanation has not been tested further.
\item \textbf{At the floors of~\cite{alrawachy2025} this device gives $K_{\rm eff}=2$, $r_{\max}=1$.}
      With a $-60$\,dBc floor and the $3$\,dB margin of Section~\ref{sec:apullsim}, five
      pairs are detected at each drive, and they are not the same five: $(0,0)$,
      $(1,\pm1)$, $(2,0)$ and $(2,2)$ at the softer drive, and $(0,0)$, $(1,\pm1)$,
      $(2,0)$ and $(2,-2)$ at the harder, where $(2,2)$ at $-57.2$\,dBc misses the margin
      by $0.2$\,dB. All have $r\le1$ and none exceeds $m=2$. The
      default ASM-HEMT has a weaker load-side nonlinearity than the device in
      \cite{alrawachy2025}, whose $m=3$ terms lie near $-35$ to $-45$\,dBc; here they
      are at $-68$\,dBc and below, so the simulation does not reproduce the measured set
      of pairs.
\item \textbf{$(4,0)$ is the only $r=2$ term above the floor at either drive}:
      $-71.8$\,dBc at the softer drive, $8$\,dB above the $|n|\le8$ level and equal to the
      $|n|\le6$ level, and $-97.6$\,dBc at the harder, $9$\,dB above the $|n|\le8$ level
      and below the $|n|\le6$ level. The other $r=2$ lines are below the floor, so the
      simulation is consistent with the ordering argued in Section~\ref{sec:apull} and
      cannot confirm it. Among the lines with $m\ge4$ that are above the floor, $(4,0)$ is
      the strongest at the softer drive, while at the harder drive the $r=1$ term
      $(4,-2)$ exceeds it --- the interleaving that Section~\ref{sec:apullsim} also found.
\item \textbf{The decay in $r$ is not geometric.}
      Along $n=0$ the first two steps are $-44$ and $-28$\,dB at the softer drive and
      $-52$ and $-46$\,dB at the harder. A one-step extrapolation from the first ratio, which is
      what Section~\ref{sec:apull} did with the measured coefficients, would put $(4,0)$
      at $-88$\,dBc and $-104$\,dBc; the model puts it at $-72$\,dBc and $-98$\,dBc. So that
      extrapolation is low by $16$\,dB in one case and $6$\,dB in the other, both figures
      resting on lines within $10$\,dB of the floor.
\end{itemize}

One incident during this run is worth recording because it illustrates the diagnostic. A sign error in the phase reference of the phasor extraction imposed a
run-to-run phase jitter of a few milliradians on every point. The effect on the spectrum
was a flat floor of lines at $-55$ to $-70$\,dBc at every bin, on the cluster positions
and in the empty bins between them, under which every $m\ge3$ term was buried: the spurious floor of a phase-referenced
measurement. The diagnostic that caught it is
the one \texttt{xp\_apull\_read} now reports first: the level of the bins that belong to no
cluster of low order. Those bins must be empty. If they are not, the data has a floor of its own, and
nothing below it should be read as device response.

Finally, the DC drain current $i_{2,0}$ was read off in the same way. Its clusters at $n$
and $-n$ have identical levels, $(1,\pm1)$ at $-29.0$\,dBc, $(2,\pm2)$ at $-62.2$\,dBc, which is the
statement that a weight-zero equivariant of a real quantity has conjugate-symmetric
coefficients --- a small check that the bookkeeping of Section~\ref{sec:equivariance} is
being applied to the right object.

\section{Summary of the identification procedure}

The procedure suggested here for identifying a Cardiff model is summarized as follows. It has not been tested on measured data.

\begin{enumerate}[leftmargin=2em]
\item \textbf{Convert your Cardiff exponents to $(a,b)$.} $a=(m+n)/2$, $b=(m-n)/2$. Then
      $r=\min(a,b)$, and the rest of this list applies.
\item \textbf{Interpret $m$ as the order of the load-side nonlinearity.} The largest $m$ that can be resolved is the
      order of the device's load-side nonlinearity at that drive and mismatch.
\item \textbf{Treat $r\le1$ as a hypothesis about the device.} It is exact if $K\le3$.
      If $r=2$ terms are needed to fit the data, the device has fourth-order load-side
      content.
\item \textbf{Plot coefficients against $r$ on a log axis.} For a loaded device the
      truncation point is where that line meets the floor. The slope reflects the size of
      the measured region compared with the radius of convergence of the series, and is
      worth reporting.
\item \textbf{If you have A-pull apparatus, measure the cluster widths.} The cluster widths give $K$
      by inspection of the spectrum rather than by a fit. Simulate the floor sweep with \texttt{xp\_apull\_demo} first.
\item \textbf{Distinguish two kinds of residual.} One that shrinks as terms are added is
      truncation, which for a loaded device is always present. One that stays at the same
      level as terms are added lies outside the form: the representation
      \eqref{eq:complete} is complete for time-invariant,
      single-tone, steady-state operation, so such a residual means another tone, another port,
      harmonic injection, or a measurement that is not in steady state. Memory shows up
      under modulation, not as a missing CW term.
\end{enumerate}

\section{Scope}

The mathematics used here is classical. The invariant and equivariant structure of a
circle action is textbook material; the standard reference is Golubitsky, Stewart and
Schaeffer~\cite[Ch.~XII and~XVI]{gss1988}. Schwarz's theorem~\cite{schwarz1975} is the
smooth extension, needed only at zero load wave. The completeness statement is a translation of the form Tasker and
Benedikt published~\cite{tasker2011,tasker2009dml} and the 2010 Cardiff paper fitted~\cite{woodington2010}; the
proof of the symmetry from a time shift, for one input and with memory, is
Enzinger's~\cite{enzinger2018}; the monomial rewriting is Tian's~\cite{tian2025}; and the
counting of which monomials reach which harmonic is the frequency-mix bookkeeping of
weakly nonlinear analysis~\cite{bussgang1974}.

What is offered here is the translation itself, the identification of $r=\min(a,b)$ as a count
of conjugate pairs in Tian's rewriting, the bound it implies, the interpretation of the
A-pull spectrum that follows, and two simulations of the proposed measurement, one on a
polynomial device and one on a compact transistor model. None of these statements has been
tested against a measurement in which $K$ is known independently.

The scope is narrow: a two-port device in fundamental
load-pull under single-tone CW excitation, in periodic steady state, and, for
Theorem~\ref{thm:rmax}, with a memoryless load-side nonlinearity. The multi-harmonic form of the Cardiff model, with one phase
exponent per harmonic, is written out by Tasker and Benedikt~\cite{tasker2011} and in
Bell's thesis~\cite{bell2014}; the multi-harmonic and multi-port cases use the same
circle, with more weights and more indices to track; only incommensurate tones, each with
its own phase origin, call for a torus in place of the circle, as in
Section~\ref{sec:symmetry}. Memory lies
outside this tutorial. It is the subject of the second companion paper~\cite{paper2},
which reads the long-term memory of a device at a large-signal operating point as
motion on a slow invariant manifold whose dimension is the
number of slow Floquet exponents, and works the construction out in simulation for a
transistor model with thermal and trap memory. The circle-group material of this tutorial
is the background that paper assumes.

\section{Glossary}
\label{sec:glossary}

\begin{center}
\begin{tabular}{p{0.20\textwidth}p{0.36\textwidth}p{0.36\textwidth}}
\toprule
term & what it means & where it appeared here \\
\midrule
group &
transformations that compose, undo, and include the identity &
time shifts; sign flips \\[2pt]
Lie group &
a group with continuous parameters, smoothly composed &
$U(1)$, the circle of phase shifts \\[2pt]
$U(1)$, $S^1$, $SO(2)$ &
three names for the unit circle &
the phase knob \\[2pt]
torus $T^{N}$ &
several independent circles &
one phase origin per incommensurate tone \\[2pt]
action &
the rule saying how each quantity transforms &
$A_{p,k}\mapsto A_{p,k}e^{jk\theta}$ \\[2pt]
representation &
the collected actions on all the variables &
the assignment of weights to harmonics \\[2pt]
weight &
how fast a quantity rotates (a physicist's charge) &
the harmonic index $k$ \\[2pt]
orbit &
the set swept out by one quantity under the group &
a circle of constant $|A|$ \\[2pt]
invariant &
unchanged by every group element &
$|\Aone|$, $|\Atwo|$, relative phase $Q$ \\[2pt]
equivariant &
transforms with a prescribed weight &
the response $B_{p,h}$, weight $h$ \\[2pt]
generators &
the finite list everything else is built from &
two magnitudes and one relative phase \\[2pt]
gauge fixing &
choosing a convention that cancels out &
the carrier $P=e^{j\phi_{1,1}}$ \\
\bottomrule
\end{tabular}
\end{center}

\appendix

\section{The bound on $r$ without group theory}
\label{app:noalgebra}

For a reader who wants only the bound, the argument follows with the group-theoretic
vocabulary removed.

A device whose load-side nonlinearity is a polynomial of degree $K$ produces, at most,
terms in which the load-side wave appears $K$ times. Each appearance is either $A$ or
$\Ab$. Write a term as $A^{a}\Ab^{\,b}$; then $a+b\le K$.

The Cardiff model writes the same term as $|A|^{m}(\angle A)^{n}$ with $m=|n|+2r$. Matching
the two gives $m=a+b$, $n=a-b$ and $r=\min(a,b)$.

A term with $r$ conjugate pairs needs at least $2r$ factors, so $2r\le a+b\le K$, hence
\[
r \;\le\; \lfloor K/2 \rfloor .
\]
For $K=3$ this gives $r\le1$: the familiar restriction, exactly. For $K=4$ it gives
$r\le2$, and the newly available term is $A^2\Ab^{\,2}=|A|^4$.

\section{Companion MATLAB routines}
\label{app:matlab}

The routines listed below, the two ngspice ASM-HEMT datasets of
Section~\ref{sec:asmhemt}, and the scripts that produced them are archived as a
versioned toolkit with a DOI~\cite{toolkit}. The routines listed need base MATLAB only;
\texttt{loaded\_exact} uses the Parallel Computing and Symbolic Math toolboxes when they
are present and runs without them; \texttt{xp\_selftest} checks every routine against values computed
independently and should be run first. The \texttt{matlab/} and \texttt{spice/} folders
of the archive must be kept side by side, since \texttt{xp\_apull\_asmhemt} locates the
datasets relative to its own location.

\begin{center}
\begin{tabular}{lp{0.62\textwidth}}
\toprule
routine & what it does \\
\midrule
\texttt{xp\_monomials} & lists monomials $A^{a}\Ab^{b}$ of given total degree and weight;
the lattice of Figure~\ref{fig:lattice} \\
\texttt{xp\_cardiff\_count} & counts what $r\le1$ keeps and discards at each degree; the
table in Section~\ref{sec:monomials} \\
\texttt{xp\_rmax\_rule} & expands a degree-$K$ device, selects harmonics, tabulates the
observed $\max\min(a,b)$ against $\lfloor K/2\rfloor$ \\
\texttt{xp\_device\_demo} & the same sweep framed as the cost of the restriction \\
\texttt{xp\_molien}, \texttt{xp\_series} & Molien series for a weighted circle action:
how many invariants and equivariants at each degree \\
\texttt{xp\_design}, \texttt{xp\_completeness} & build the design matrix of weight-$h$
monomials at sample points and compare its numerical rank with the Molien dimension:
completeness shown as a rank rather than quoted as a theorem \\
\texttt{xp\_radial\_demo} & in one load variable, measures the residual left by the
$r\le1$ truncation as the degree grows: what the restriction costs \\
\texttt{xp\_loaded\_demo} & closes the load loop on a quadratic device, fits the monomials
over a disc and reports the largest coefficient in each $r$-class; the open-loop row of
the table in Section~\ref{sec:loaded} \\
\texttt{loaded\_exact} & the exact series of the same loaded device in a reference
impedance $Z$: coefficients, radius of convergence, truncation error; the closed-loop
rows and the numbers of Section~\ref{sec:loaded} \\
\texttt{xp\_apull\_demo} & tailored A-pull identification on a synthetic Curtice-type
device: spiral stimulus, FFT over the sequence, exponent pairs read off the clusters,
floor sweep; Figure~\ref{fig:apullsim} and the table of Section~\ref{sec:apullsim} \\
\texttt{xp\_apull\_read} & the analysis half of the above, for any measured or simulated
response sequence; reports the off-cluster floor first \\
\texttt{xp\_apull\_asmhemt} & loads the ngspice ASM-HEMT datasets in \texttt{spice/} and
reads them; Figure~\ref{fig:asmhemt} and Table~\ref{tab:asmhemt} \\
\texttt{xp\_selftest} & runs every check above against values computed independently \\
\bottomrule
\end{tabular}
\end{center}

\begin{lstlisting}
>> xp_selftest                     % every routine against known values
>> T = xp_rmax_rule(2:7);          % Section 7
>> T = xp_cardiff_count([3 4 5 9 12]);   % Section 6
>> xp_loaded_demo(0.5, false);     % open loop: r_max = 1 for K = 2
>> xp_loaded_demo(0.5);            % closed loop: every r present, decaying
>> loaded_exact                    % exact series, radius of convergence
>> xp_apull_demo(3, true, -60);    % Section 10.7: seven pairs, all r <= 1
>> xp_apull_demo(3, true, -70);    % ... and (4,0) appears
>> xp_apull_asmhemt(-60);          % Section 10.8: the ASM-HEMT data
\end{lstlisting}

\section*{Acknowledgment}

The author thanks the two anonymous reviewers of the companion paper~\cite{paper1}, whose
reports identified errors that are corrected in this tutorial as well. The author used
Claude (Anthropic) to edit for clarity, check algebra, and assist with the figures and
simulation scripts; the author reviewed and validated all AI-assisted output and takes
full responsibility for the content.



\begin{thebibliography}{99}
\bibitem{paper1}
N.~B. Tufillaro, ``Time invariance, circle symmetry, and the completeness of the Cardiff
behavioral model,'' arXiv:2609.35828 [eess.SP], 2026.

\bibitem{toolkit}
N.~B. Tufillaro, ``xpar-toolkit: companion MATLAB and ngspice tools for `Symmetry and the Form
of Nonlinear Behavioral Models','' version 1.2.0, Zenodo, 2026, doi:10.5281/zenodo.24602291.

\bibitem{qi2009}
H.~Qi, J.~Benedikt, and P.~J. Tasker, ``Nonlinear data utilization: From direct data
lookup to behavioral modeling,'' \emph{IEEE Trans. Microw. Theory Techn.}, vol.~57,
no.~6, pp.~1425--1432, 2009.

\bibitem{woodington2010}
S.~Woodington, R.~Saini, D.~Williams, J.~Lees, J.~Benedikt, and P.~J.~Tasker,
``Behavioral model analysis of active harmonic load-pull measurements,''
\emph{IEEE MTT-S Int. Microwave Symp. Dig.}, 2010, pp.~1688--1691.
\textit{The traveling-wave Cardiff form, and the observation that the device responds only
to magnitudes and relative phase.}

\bibitem{tasker2011}
P.~J. Tasker and J.~Benedikt, ``Waveform inspired models and the harmonic balance
emulator,'' \emph{IEEE Microw. Mag.}, vol.~12, no.~2, pp.~38--54, 2011.

\bibitem{tasker2009dml}
P.~J. Tasker, ``The large signal microwave characterization and design challenge,'' IEEE
MTT-S Distinguished Microwave Lecture, Trondheim, Norway, 2009, slides 28--34. [Online].
Available: https://ewh.ieee.org/r8/norway/ap-mtt/files/Tasker\_DML\_Trondheim2009.pdf
\textit{The general form $b=P\cdot f(|a_1|,|a_2|,Q/P)$ with its Fourier expansion in $Q/P$.}

\bibitem{root2005phd}
D.~E. Root, J.~Verspecht, D.~Sharrit, J.~Wood, and A.~Cognata, ``Broad-band
poly-harmonic distortion (PHD) behavioral models from fast automated simulations and
large-signal vectorial network measurements,'' \emph{IEEE Trans. Microw. Theory Techn.},
vol.~53, no.~11, pp.~3656--3664, 2005.
\textit{The PHD model, and the statement that the phase factor and the magnitude-only
coefficients follow from time invariance.}

\bibitem{root2013}
D.~E.~Root, J.~Verspecht, J.~Horn, and M.~Marcu,
\emph{X-Parameters}. Cambridge University Press, 2013.
\textit{Where the phase-normalization factor $P^{k}$ appears; Section~\ref{sec:equivariance}
identifies it as an equivariant carrier.}

\bibitem{nuij2006}
P.~W.~J.~M.~Nuij, O.~H.~Bosgra, and M.~Steinbuch,
``Higher-order sinusoidal input describing functions for the analysis of non-linear systems
with harmonic responses,''
\emph{Mech. Syst. Signal Process.}, vol.~20, no.~8, pp.~1883--1904, 2006.
\textit{The same construction under the name ``virtual harmonics generator''.}

\bibitem{enzinger2018}
H.~Enzinger, ``Behavioral modeling and digital predistortion of radio frequency power
amplifiers,'' Dr.~techn. thesis, Graz Univ. of Technology, Graz, Austria, 2018.
\textit{Secs.~3.2.7 and~3.3.5: ``phase homogeneity,'' the single-input form of the
circle action, proved for a general time-invariant operator with memory.}

\bibitem{paper2}
N.~B. Tufillaro, ``Memory in behavioral models as motion on a slow invariant manifold,''
arXiv:2609.38771 [eess.SP], 2026.

\bibitem{gibiino2020}
G.~P. Gibiino, A.~M. Angelotti, A.~Santarelli, F.~Filicori, and P.~A. Traverso, ``M$^2$S
parameters: A multi-tone multi-harmonic measurement approach for the characterization of
nonlinear networks,'' in \emph{Proc. IEEE Int. Instrum. Meas. Technol. Conf. (I2MTC)},
2020, pp.~1--6.
\textit{Two incommensurate tones; the two-torus case of Section~\ref{sec:symmetry} in
measurement practice.}

\bibitem{kurokawa1965}
K.~Kurokawa,
``Power waves and the scattering matrix,''
\emph{IEEE Trans. Microw. Theory Techn.}, vol.~13, no.~2, pp.~194--202, 1965.
\textit{Reference-plane phase conventions; the extra circles of
Section~\ref{sec:symmetry}.}

\bibitem{schwarz1975}
G.~W.~Schwarz,
``Smooth functions invariant under the action of a compact Lie group,''
\emph{Topology}, vol.~14, pp.~63--68, 1975.
\textit{The step from polynomial to smooth; why the completeness claim applies to real
devices.}

\bibitem{tian2025}
M.~Tian, J.~J. Bell, R.~Quaglia, E.~M. Azad, and P.~J. Tasker, ``Artificial neural network
nonlinear transistor behavioral models: Structure and parameter determination process
based on the Cardiff model,'' \emph{IEEE Trans. Microw. Theory Techn.}, vol.~73, no.~2,
pp.~745--759, Feb. 2025.
\textit{Appendix~A rewrites the Cardiff model as a polynomial in $A_{2,1}$ and
$\bar A_{2,1}$; Section~\ref{sec:monomials} starts from that rewriting.}

\bibitem{ding2004}
L.~Ding and G.~T. Zhou, ``Effects of even-order nonlinear terms on power amplifier
modeling and predistortion linearization,'' \emph{IEEE Trans. Veh. Technol.}, vol.~53,
no.~1, pp.~156--162, Jan. 2004.

\bibitem{demellokoch2026}
R.~de~Mello~Koch, M.~Kim, A.~L.~Mahu, and A.~Rudra,
``A pedagogical introduction to invariant theory and finite-$N$ holography,''
arXiv:2608.22952, 2026.
\textit{Hilbert series, Molien--Weyl and the primary/secondary decomposition, explained to
people who know vector spaces; Sections~2.3--2.8.}

\bibitem{weyl1936}
H.~Weyl,
\emph{Elementary Theory of Invariants}, lecture notes, Institute for Advanced Study,
Princeton, 1935--36 (notes by H.~Weyl and L.~M.~Blumenthal).
\textit{The finiteness theorems in Weyl's own lectures, three years before}
The Classical Groups.

\bibitem{verspecht2023dg}
J.~Verspecht, S.~Kusano, and A.~Stav, ``The dynamic gain model,'' in \emph{Proc. 100th
ARFTG Microw. Meas. Conf.}, 2023, pp.~1--4.

\bibitem{azad2022}
E.~M.~Azad, J.~J.~Bell, R.~Quaglia, J.~J.~Moreno Rubio, and P.~J.~Tasker,
``New formulation of Cardiff behavioral model including DC bias voltage dependence,''
\emph{IEEE Microw. Wireless Compon. Lett.}, vol.~32, no.~6, pp.~607--610, 2022.

\bibitem{alrawachy2025}
A.~Al-Rawachy, A.~Baddeley, A.~Eblabla, D.~Gecan, A.~Sheikh, A.~Bogusz, R.~Quaglia, and
P.~J.~Tasker,
``Automated Cardiff model complexity identification and parameters extraction from measured
tailored A-pull data,''
\emph{IEEE J. Microwaves}, vol.~5, no.~5, pp.~1150--1161, 2025.
\textit{The measurement discussed in Section~\ref{sec:apull}.}

\bibitem{tasker2020}
P.~J.~Tasker,
``Robust extraction of Cardiff model parameters from appropriately tailored measured
load-pull data,''
\emph{Proc. IEEE BiCMOS Compound Semicond. Integr. Circuits Technol. Symp.}, 2020,
pp.~1--5.

\bibitem{khandelwal2018asmhemt}
S.~Khandelwal, S.~Ghosh, S.~A.~Ahsan, A.~Dasgupta, and Y.~S.~Chauhan,
``ASM-HEMT 101.0.0: Advanced SPICE model for HEMTs --- technical manual,''
IIT Kanpur and Macquarie Univ., 2018.
\textit{The compact model used in Section~\ref{sec:asmhemt}.}

\bibitem{gss1988}
M.~Golubitsky, I.~Stewart, and D.~G.~Schaeffer,
\emph{Singularities and Groups in Bifurcation Theory}, Vol.~II. Springer, 1988.
\textit{Chapter~XII is the standard treatment; the weighted circle and torus cases appear
there as worked exercises.}

\bibitem{bussgang1974}
J.~J. Bussgang, L.~Ehrman, and J.~W. Graham, ``Analysis of nonlinear systems with multiple
inputs,'' \emph{Proc. IEEE}, vol.~62, no.~8, pp.~1088--1119, Aug. 1974.

\bibitem{bell2014}
J.~J.~W. Bell, ``Input harmonic and mixing behavioural model analysis,'' Ph.D. thesis,
Cardiff Univ., Cardiff, U.K., 2014.
\textit{The multi-harmonic Cardiff formulation, one phase exponent per harmonic.}

\end{thebibliography}
\end{document}